\documentclass[10pt]{article}
\usepackage{amssymb}
\usepackage{graphicx}

\font\gg=eufm10 at 12pt
\def\goth#1{\hbox{\gg #1}}
\def\bthm#1{\vskip10pt\noindent{\bf#1}\space\bgroup\em}
\def\ethm{\egroup\vskip10pt}
\def\rem{\vskip10pt\noindent}
\def\bbox{\mbox{\large\lower.3ex\hbox{\large$\Box$}}}
\def\forall{\hbox{ for all }}
\let\dsty\displaystyle

\begin{document}

\def\Span{\mathop{\rm span}\nolimits}
\def\qed{\hfill$\Box$\ \par}

\vspace* {2mm}
\begin{center}
{\bf STUDY OF GRAM MATRICES IN FOCK REPRESENTATION OF MULTIPARAMETRIC
CANONICAL COMMUTATION RELATIONS, EXTENDED ZAGIER'S CONJECTURE,
HYPERPLANE ARRANGEMENTS AND QUANTUM GROUPS\footnote{Published in Math.\ Commun.\ {\bf 1}, 1--24, (1996)}
.}
\\[1ex]
S. Meljanac$^1$ and D. Svrtan$^2$ \\[1mm]
{\it $^1$ Rudjer Bo\v skovi\'c Institute - Bijeni\v cka c. 54, 10000 Zagreb,
Croatia \\[1mm]
$^2$ Dept. of Math., Univ. of Zagreb, Bijeni\v cka c. 30, 10000 Zagreb, Croatia}
\end{center}
{\bf Abstract}.In this Colloqium Lecture (by one of the authors (D.S))
a thorough presentation of the authors research on the subjects ,stated in the
title,is given.By quite laborious mathematics it is explained how one can
handle systems in which each Heisenberg commutation relation is deformed
separately.For Hilbert space realizability a detailed determinant computations
(extending Zagier's one parametric formulas) are carried out.The inversion
problem of the associated Gram matrices on Fock weight spaces is completely
solved (Extended Zagier's conjecture) and a counterexample to the original
Zagier's conjecture is presented in detail.\\[1ex]
{\bf Sa\v{z}etak}.U ovom Kolokviju (jednog od autora (D.S)) cjelovito su prikazana
i\-stra\-\v{z}i\-va\-nja autora o temama formuliranima u naslovu.S poprili\v{c}no matematike
obja\v{s}njeno je kako se mogu obradivati sustavi u kojima je svaka Heisenbergova
komutacijska relacija deformirana odvojeno.Za realizabilnost na Hilbertovu
prostoru provedeno je detaljno ra\v{c}unanje determinanata (koje pro\v{s}iruje
Zagierove jednoparametarske formule).Problem inverzije pridru\v{z}enih Gramovih
matrica na Fockovim te\v zinskim prostorima je potpuno rije\v{s}en (Pro\v {s}irena
Zagierova hipoteza) i kontraprimjer za originalnu Zagierovu hipotezu je detaljno
prikazan.\\[1ex]
{\it Klju\v{c}ne rije\v{c}i i pojmovi}.Multiparametarske kanonske komutacijske relacije,deformirane
parcijalne derivacije,re\v{s}etka subdivizija,deformirana regularna reprezentacija,
kvantna bilinearna forma,Zagierova hipoteza.\\[3mm]

 \begin{center}
{\Large\bf Introduction}\addcontentsline{toc}{section}{Introduction}
\end{center}
Following Greenberg, Zagier, Bo\v zejko and Speicher and others we study
a collections of operators $a(k)$ satisfying the "$q_{kl}$-canonical
commutation relations "
$$a(k)a^{\dagger}(l) - q_{kl}a^{\dagger}(l)a(k) = \delta_{kl}$$
(corresponding for $q_{kl}=q$ to Greenberg (infinite) statistics, for
$q=\pm 1$ to classical Bose and Fermi statistics). We show that $n!\times n!$
matrices $A_{n}(\{ q_{kl}\} )$ representing the scalar products of $n$-particle states is positive definite for all $n$ if $|q_{kl}|<1$, all $k$,$l$,so that the above commutation relations have a Hilbert space realization in this case. This is
achieved by explicit factorizations of $A_{n}(\{ q_{kl}\} )$ as a product of
matrices of the form $(1-QT)^{\pm 1}$, where Q is a diagonal matrix and T is
a regular representation of a cyclic matrix. From such factorizations we
obtain in THEOREM 1.9.2[DETERMINANT FORMULA] explicit formulas for the determinant of $A_{n}(\{ q_{kl}\} )$ in the
generic case (which generalizes Zagier's 1-parametric formula). The problem of
computing the inverse of $A_{n}(\{ q_{kl}\} )$ in its original form is
computationally intractable (for $n=4$ one has to invert a $24\times 24$
symbolic matrix). Fortunately, by using another approach (originated by
Bo\v zejko and Speicher ) we obtain in Theorem 2.2.6 a definite answer to that inversion problem
in terms of maximal chains in so called subdivision lattices. Our algorithm in Proposition 2.2.15
for computing the entries of the inverse of $A_{n}(\{ q_{kl}\} )$ is very efficient. In particular
for $n=8$, when all $q_{kl}=q$, we found a counterexample to Zagier's
conjecture
concerning the form of the denominators of the entries in the inverse of
$A_{n}(q)$. In Corollary 2.2.8 we formulate and prove Extended Zagier's
Conjecture which turns to be the best possible in the multiparametric case
and which implies in one parametric case an interesting extension of the
original Zagier's Conjecture.By using a faster algorithm in Proposition 2.2.16 we obtain in THEOREM
2.2.17[INVERSE MATRIX ENTRIES] explicit formulas for the inverse of the matrices $A_n(\{q_{kl}\})$ in
the generic case.
Finally, there are applications of the results above to discriminant
arrangements of hyperplanes and to contravariant forms of certain quantum
groups. \vskip1pt

\pagestyle{plain}
\def\leer{\vspace{2mm}}
\setcounter{equation}{0}
\section{Multiparametric quon algebras, Fock-like representation and
determinants} 
\subsection{$q_{ij}$-canonical commutation relations} 
Let ${\bf q} = \{ q_{ij} : i,j \in I,\bar{q}_{ij}=q_{ji} \} $ be a
 hermitian family of complex numbers (parameters), where I is a finite
(or infinite) set of indices.Then by a {\em multiparametric quon algebra} ${\cal A} ={\cal A}^{({\bf q})}$ we shall mean
an associative (complex) algebra generated by $\{ a_{i}, a_{i}^{\dagger},
i \in I \}$ subject to the following $q_{ij}$- canonical commutation relations
$$a_{i}a_{j}^{\dagger} = q_{ij}a_{j}^{\dagger}a_{i} + \delta_{ij},\ \
 \ \forall i,j \in I.$$
Shortly, we shall give an explicit Fock-like representation of the algebra
${\cal A}^{({\bf q})}$ on the free associative algebra ${\bf f}$
(the algebra of
noncommuting polynomials in the indeterminates $\theta _{i}, i \in I$)
with $a_{i}$ acting
as a generalized $q_{ij}$-deformed partial derivatives ${}_{i}\partial =
{}_{i}^{{\bf q}}\partial$ w.r.t. variable $\theta _{i}$ (the i-th annihilation operator),
and
$a_{i}^{\dagger}$ as multiplication by $\theta _{i}$
(the i-th creation operator). Moreover $a_{i}^{\dagger}$ will be adjoint to
$a_{i}$ w.r.t. certain sesquilinear form $(\ ,\ )_{{\bf q}}$
on ${\bf f}$ which will be better described via certain canonical ${\bf q}$-deformed bialgebra
structure on ${\bf f}$, generalizing the one used by Lusztig in his excellent
treatment of quantum groups [Lus]. Then by explicit computation (which extends
Zagier's method) of the determinant of $(\ ,\ )_{{\bf q}}$ we show that
$(\ ,\ )_{{\bf q}}$
is positive definite provided the following condition on the parameters
$q_{ij}$ holds true: $ |q_{ij}| < 1,\ \ \forall i,j \in I$
    This condition ensures that all the many-particle states
$a_{i_{1}}^{\dagger}\cdots a_{i_{r}}^{\dagger}|0> = \theta _{i_{1}}
\cdots \theta _{i_{r}}$,\ $i_{j} \in I, r \geq 0$,
are linearly independent, so we obtain a Hilbert space realization
of the $q_{ij}$-canonical commutation relations.
We first need some notations:
\begin{description}
\item[\hbox{${\bf N}=$}] $\{ 0,1,2,\dots \} =$ the set of nonnegative integers ,${\bf C}=$ the set of complex numbers
\item[\hbox{$({\bf N}[I],+) =$}] the {\em weight monoid} i.e. the set of all
         finite formal linear     combinations $\nu = \sum _{i \in I}\nu _{i}i$,
        $\nu _{i}\in {\bf N}, i\in I$ with componentwise addition
        $\nu + \nu^{'} = \sum_{i\in I}( \nu _{i} + \nu _{i} ^{'})i$
\item[$|\nu |=$] $\sum_{i\in I}\nu _{i}  \in {\bf N}$ for $\nu = \sum_{i\in I}
        \nu _{i}i  \in {\bf N}[I]$
\item[$\beta :$] $({\bf N}[I],+)\times ({\bf N}[I],+) \longrightarrow ({\bf C},
     \cdot )$,
        the bilinear form on $({\bf N}[I],+)$
    given by $i,j \mapsto q_{ij}$, i.e.
    for $\nu = \sum_{i\in I}\nu _{i}i$, $\nu^{'} = \sum_{j\in I}
    \nu _{j}^{'}j$, $\beta (\nu ,\nu^{'}) = \prod_{ij}q_{ij}
    ^{\nu _{i}\nu _{j}^{'}}$.
\end{description}

\subsection{The algebra ${\bf f}$ } 

We denote by ${\bf f}$ the free associative ${\bf C}$-algebra with generators
$\theta _{i}(i\in I)$. For any weight $\nu = \sum_{i\in I}\nu _{i}i \in
{\bf N}[I]$ we denote by ${\bf f}_{\nu }$ the corresponding {\em weight space},
i.e. the subspace of ${\bf f}$ spanned by monomials
$\theta _{{\bf i}} = \theta _{i_{1}}\cdots \theta _{i_{n}}$ indexed by
sequences ${\bf i}=i_{1}\dots i_{n}$ of weight $\nu $,$|{\bf i}|=\nu $
(this means that the number of occurrences of $i$ in ${\bf i}$ is equal
to $\nu _{i}, \forall i \in I$). Then each ${\bf f}_{\nu }$ is a finite
dimensional complex vector space and we have a direct sum decomposition
${\bf f} = \bigoplus
_{\nu }{\bf f}_{\nu }$, where $\nu $ runs over ${\bf N}[I]$. We have
${\bf f}
_{\nu }{\bf f}_{\nu ^{'}}
\subset {\bf f}_{\nu +\nu ^{'}}$, $1\in {\bf f}_{0}$ and $\theta _{i}\in
{\bf f}_{(i)}$.
An element $x$ of ${\bf f}$ is said to be {\em homogeneous} if it belongs to
${\bf f}_{\nu }$
for some $\nu $. We than say that $x$ has {\em weight} $\nu $ and write $|x|=\nu $.\\
We consider the tensor product ${\bf f}\otimes {\bf f}$ with the
following $q_{ij}$-{\em deformed multiplication}
$$
(x_{1}\otimes x_{2})(x_{1}^{'}\otimes
x_{2}^{'})=
(\prod_{i,j}q_{ij}^{\nu_{i}\nu_{j}^{'}})x_{1}x_{1}^{'}\otimes
x_{2}x_{2}^{'}, \hbox{ if }x_{2}\in {\bf f}_{\nu }, \ x_{1}^{'}\in
{\bf f}_{\nu^{'}}
$$
where $x_{1}, x_{1}^{'}, x_{2}, x_{2}^{'}\in
{\bf f}$ are homogeneous; this algebra is associative since $\beta
(\nu ,\nu^{'})$ is bilinear. The following statement is easily
verified: if $r=r_{{\bf q}} : {\bf f} \longrightarrow {\bf
f}\otimes {\bf f}$ is the unique algebra homomorphism such that
$r(\theta _{i} )=\theta_{i}\otimes 1+1\otimes \theta_{i}, \forall
i$, then
\begin{eqnarray*}
r(\theta_{i}\theta_{j})&=&r(\theta_{i})r(\theta_{j})
        \; =\; 
             \theta_{i}\theta_{j}\otimes 1+q_{ij}\theta_{j}\otimes \theta_{i}+
\theta_{i}\otimes \theta_{j}+1\otimes \theta_{i}\theta_{j}
\end{eqnarray*}
\noindent
More generally, the value of $r$ on any monomial $\theta_{{\bf
i}}=\theta_{i_{1}}\theta_{i_{2}} \cdots \theta_{i_{n}}$\, is given by:
$$r(\theta_{{\bf i}})=\sum_{k+l=n, g=(k,l)-shuffle}q_{{\bf i},g}
\theta_{i_{g(1)}}\cdots \theta_{i_{g(k)}}\otimes \theta_{i_{g(k+1)}}
\cdots \theta_{i_{g(k+l)}}$$ where $(k,l)-shuffle$ is a permutation
$g\in S_{k+l}$ such that $g(1)<g(2)<\cdots <g(k)$ and
$g(k+1)<g(k+2)<\cdots <g(k+l)$ and where for $g\in S_{n}$ we denote by
$q_{{\bf i},g}$the quantity $$q_{{\bf
i},g}:=\prod_{a<b,g(a)>g(b)}q_{i_{a}i_{b}}$$

\subsection{The sesquilinear form $(\ ,\ )_{{\bf q}}$ on ${\bf f}$ } 
 Note that $r$ maps ${\bf f}_{\nu }$ into $\bigoplus_{(\nu^{'}+\nu^{''}=\nu )}
{\bf f}_{\nu^{'}}\bigotimes {\bf f}_{\nu^{''}}$. Then the linear maps
${\bf f}_{\nu^{'}+\nu^{''}}\longrightarrow {\bf f}_{\nu^{'}}\bigotimes
{\bf f}_{\nu^{''}}$ defined by $r$ give,
by passage to dual spaces, linear maps ${\bf f}_{\nu^{'}}^{*}\bigotimes
{\bf f}_{\nu^{''}}^{*}\longrightarrow {\bf f}_{\nu^{'}+\nu^{''}}^{*}$.
These define the structure of an associative algebra with 1 on $\bigoplus
_{\nu }{\bf f}_{\nu }^{*}$.
For any $i\in I$, let $\theta_{i}^{*}\in {\bf f}_{i}^{*}$ be the linear
form given by $\theta_{i}^{*}(\theta_{j})=\delta_{ij}$. Let ${\Phi}_{\bf q} :
{\bf f}\longrightarrow \bigoplus_{\nu }{\bf f}_{\nu }^{*}$ be the unique
conjugate-linear algebra homomorphism preserving 1, such that
${\Phi}_{\bf q} (\theta_{i})=
\theta_{i}^{*},\forall i$.
For $x,y \in {\bf f}$, we set
$$(x,y)_{{\bf q}} = {\Phi}_{\bf q} (y)(x)$$
Then $(\ ,\ )=(\ ,\ )_{{\bf q}}$ is a unique sesquilinear form on ${\bf f}$
such that {\bf a)} $(\theta_{i},\theta_{j})=\delta_{ij}, \  \forall i,j \in I$;
{\bf b)}  $(x,y^{'}y^{''})=(r(x),y^{'}\bigotimes y^{''}), \  \forall x,y^{'},y^{''}
\in {\bf f}$; \,{\bf c)} $(xx^{'},y^{''})=(x\bigotimes x^{'},r(y^{''})), \
\forall x,x^{'},y^{''} \in {\bf f}$.Clearly,{\bf d)}  $(x,y)=0$ if $x$ and $y$ are
homogeneous with $|x|\neq |y|$.Thus the subspaces ${\bf f}_{\nu }, {\bf f}_{\nu^{'}}$ are orthogonal w.r.t.
$(\ ,\ )$
        for $\nu \neq \nu^{'}$.

             \subsection{The $q_{ij}$-deformed partial derivative maps $_{i}^{{\bf q}}
\partial $ and $^{{\bf q}}\partial_{i}$ } 

Let $i\in I$. Clearly there exists a unique ${\bf C}$-linear map
${}_{i}\partial ={}_{i}^{{\bf q}}\partial :{\bf f}\longrightarrow {\bf f}$ such
that
${}_{i}\partial (1)=0$, ${}_{i}\partial (\theta_{j})=\delta_{ij}$, \ $\forall j$
and obeying the generalized Leibniz rule :\\
{\bf a)}  $_{i}\partial (xy)={}_{i}\partial (x)y + \beta (i,|x|)x _{i}\partial (y)$                 $ ={}_{i}\partial (x)y + \prod_{j}q_{ij}^{\nu_{j}}x_{i}\partial (y)$, if $x\in {\bf f}_{\nu }$\\
for all homogeneous $x$,$y$. If $x\in {\bf f}_{\nu }$ we have $_{i}
\partial (x)
\in {\bf f}_{\nu -i}$ if $\nu_{i} \geq 1$ and $_{i}\partial (x)=0$
if $\nu_{i}=0$;
 moreover
$r(x)=\theta_{i}\bigotimes {}_{i}\partial (x)$ $ +
$ terms of other bihomogeneities.
\noindent
Similarly,we define unique ${\bf C}$-linear map
$\partial_{i}={}^{{\bf q}}\partial_{i} : {\bf f}\longrightarrow {\bf f}$
such that
$\partial_{i}(1)=0$,\ $\partial_{i}(\theta_{j})=\delta_{ij}$
for all $j$ and $\partial_{i}(xy)=\beta (|y|,i)\partial_{i}(x)y +
x\partial_{i}(y)$ $(=(\prod_{j}q_{ji}^{\nu_{j}})\partial_{i}(x)y +
x\partial_{i}(y)$, if $y\in {\bf f}_{\nu }$) for all homogeneous $x$,$y$.
From the definition we see that\\
{\bf b)}  $(\theta_{i}y,x)=(y,{}_{i}\partial (x)), (y\theta_{i},x)=
(y,\partial_{i}(x))$,
        for all $x,y$;i.e. the operator ${}_{i}\partial $ (resp. $\partial_{i}$) is the adjoint of the left
(resp. right) multiplication by $\theta_{i}$.We shall need the
following explicit formula for ${}_{i}\partial ={}_{i}^
{{\bf q}}\partial :{\bf f}\longrightarrow {\bf f}$ \\
{\bf c)}  ${}_{i}\partial (\theta_{j_{1}}\cdots \theta_{j_{n}})=
\sum_{(p:j_{p}=i)}q_{ij_{1}}\cdots q_{ij_{p-1}}\theta_{j_{1}}\cdots
\hat{\theta }_{j_{p}}\cdots \theta_{j_{n}}$\\
where $\hat{}$ denotes omission of the factor $\theta_{j_{p}}$. This formula is obtained by iterating the recursive definition a) for $_{i}\partial $ or by
using the general formula for $r$ in 1.2.
Similar formula holds for $\partial_{i}$.
\subsection{Fock representation of multiparametric quon algebra
${\cal A}^{({\bf q})}$ } 
Here we give a representation of the multiparametric quon
algebra ${\cal A}={\cal A}^{({\bf q})}$ (defined in 1.1) on the underlying
vector space of the free associative algebra ${\bf f}$.
\bthm{PROPOSITION 1.5.1.} For each $i\in I$ let $a_{i}^{\dagger}$ acts on ${\bf f}$
 as left
multiplication by $\theta_{i}$ and let $a_{i}$ acts as a linear map $_{i}
\partial $ defined in 1.4. Then \\
a)  $a_{i}, a_{i}^{\dagger}$ make ${\bf f}$ into a left ${\cal A}$ - module\\
b)  $a_{i}^{\dagger}$ is adjoint to $a_{i}$ w.r.t. sesquilinear form
$(\ ,\ )=(\ ,\ )_{{\bf q}}$ defined in 1.3.\\
c)  $a_{i} : {\bf f} \longrightarrow {\bf f}$ is locally nilpotent for every
$i\in I$.
\ethm
\subsection{The matrix $A({\bf q})$ of the sesquilinear form $(\ ,\
)_{{\bf q}}$ on ${\bf f}$ } 
Here we study the sesquilinear form $(\
,\ )_{{\bf q}}$ on ${\bf f}$, defined in 1.2, via associated matrix
w.r.t. the basis $B=\{ \theta_{{\bf i}}=\theta_{i_{1}} \cdots
\theta_{i_{n}} | i_{j}\in I, n\geq 0\} $ of the complex vector space
${\bf f}=\bigoplus_{\nu }{\bf f}_{\nu }$. Let $B^{'}=\{ \theta_{{\bf
i}}= \theta_{i_{1}} \cdots \theta_{i_{n}} | i_{1},\dots ,i_{n}$ all
distinct$\} $ and $B^{''}= B\setminus B^{'}= \{ \theta_{i_{1}}\cdots
\theta_{i_{n}} |$ not all $ i_{1},\dots ,i_{n}$ distinct$\} $. Then we
have the direct sum decomposition ${\bf f} = {\bf f}^{'}\bigoplus {\bf
f}^{''}$, where ${\bf f}^{'}=\Span B^{'}$ ,  ${\bf f}^{''}=\Span B^{''}$.
Note that for any weight $\nu =\sum \nu_{i}i \in {\bf N}[I]$ we have
${\bf f}_{\nu }\subset {\bf f}^{'}$ (resp.${\bf f}_{\nu }\subset {\bf
f}^{''}$) if all $\nu_{i}\leq 1$ (resp. some $\nu_{i}\geq 2$). Then we
call such weight $\nu $ {\em generic} (resp. {\em degenerate} ) and we
have further direct sum decompositions
 ${\bf f}^{'} = {\textstyle\bigoplus}_{\nu \ generic}{\bf f}_{\nu } $,
 ${\bf f}^{''} = {\textstyle\bigoplus}_{\nu \ degenerate}
 {\bf f}_{\nu}$ .

\bthm{PROPOSITION 1.6.1.} i) Let ${\bf A}={\bf A}({\bf q}) : {\bf f}
\longrightarrow {\bf f}$ be the linear operator,
associated to the sesquilinear form $(\ , \ )=(\ ,\ )_{{\bf q}}$ on ${\bf f}$
defined by
$${\bf A}(\theta_{{\bf j}}) = \sum_{{\bf i}}(\theta_{{\bf j}},
\theta_{{\bf i}})_{{\bf q}}\theta_{{\bf i}}$$
Then the ${\bf f}^{'}, {\bf f}^{''}, {\bf f}_{\nu }\ (\nu \in {\bf N}[I])$
are all invariant subspaces
of ${\bf A}$, yielding the following block decompositions for the
corresponding matrices
\[ A=A^{'}\bigoplus A^{''},\ \
A^{'}={\bigoplus}_{\nu \ \rm generic}A^{(\nu )},\ \
 A^{''} = {\bigoplus}_{\nu \ \rm degenerate}A^{(\nu
 )},
 \]
 with entries given by the following formulas:\\

 ii) Let ${\bf i}=i_{1}\dots i_{n}$ and ${\bf j}=j_{1}\dots j_{n}$
be any two sequences with the same generic weight $\nu $ and let $\sigma =
\sigma ({\bf i,j}) \in S_{n}$ be the unique permutation
such that $\sigma \cdot {\bf i}={\bf j}$ (i.e. $i_{\sigma ^{-1}(p)}=
j_{p}$, all $p$). Then
$$A^{'}_{{\bf i,j}} = A^{(\nu )}_{{\bf i,j}} = q_{{\bf i},\sigma }
(=\bar{q}_{{\bf j},\sigma ^{-1}})$$
where (cf. 1.2) $q_{{\bf i},\sigma}:=\prod_{(a,b)\in I(\sigma)}q_{i_{a}i_{b}}$
with $I(\sigma)=\{(a,b) | a<b, \sigma (a)>\sigma (b)\}$ denoting the set
of inversions of $\sigma $.\\
 iii) Let ${\bf i}=i_{1}\dots i_{n}$ and ${\bf j}=
j_{1}\dots j_{n}$
 be any two sequences of the same degenerate weight $\nu $ and let
${ \bf \sigma  } ({\bf i,j})=\{ \sigma \in S_{n} |
 i_{\sigma^{-1} (p)}=j_{p}$, all $p\} $. Then
$$A^{''}_{{\bf i,j}}=A^{(\nu )}_{{\bf i,j}}=\sum_{\sigma \in
{ \bf \sigma  } ({\bf i,j})}q_{{\bf i},\sigma^{-1}} \ (=\sum_{\sigma \in
{ \bf \sigma  } ({\bf i,j})}\bar{q}_{{\bf j},\sigma^{-1} }).$$
\ethm
{\bf Proof.}\ i) follows from 1.3d). For ii) we have, by 1.4b)
\vskip-2ex
$$A^{'}_{{\bf i,j}}=
        A_{{\bf i,j}}=(\theta_{{\bf j}},\theta_{{\bf i}})_{{\bf q}}=
     ({}_{i_{1}}\partial (\theta_{{\bf j}}),\theta_{i_{2}}\cdots
    \theta_{i_{n}})_{{\bf q}}=\cdots ={}_{i_{n}}\partial \cdots {}_{i_{1}}
    \partial (\theta_{j_{1}}\cdots \theta_{j_{n}})$$

\noindent
By applying 1.4d) successively for $i=i_{1},i_{2},\dots $
and if $j_{\sigma (1)}=i_{1}, j_{\sigma (2)}=i_{2},\cdots $ we obtain
$(\prod_{1<b, \sigma (b)<\sigma (1)}q_{i_{1}i_{b}})(\prod_{2<b, \sigma (b)<
\sigma (2)}q_{i_{2}i_{b}})\cdots =\prod_{a<b, \sigma (b)<\sigma (a)}
q_{i_{a}i_{b}}=q_{{\bf i},\sigma },$ so the claim follows.
The proof of iii) is similar as for ii) except that
$\sigma $ is not unique.~\qed
\rem{\bf Remark 1.6.2.} For any weight $\nu =\sum \nu_{i}i$
with $|\nu |= \sum \nu_{i}=n$, the size of the matrix $A^{(\nu )}$ is
equal to $n!/{\prod_{i}\nu_{i}!}=\dim {\bf f}_{\nu }$. Hence for $\nu $
generic $A^{(\nu )}$ is an $n!\times n!$ matrix. \rem{\bf Example
1.6.3.}\ \ Let $I=\{ 1,2,3\} $ and $\nu $ generic with
$\nu_{1}=\nu_{2}=\nu_{3}=1$. Then w.r.t. basis $\{ \theta_{123},
\theta_{132}, \theta_{312}, \theta_{321}, \theta_{231}, \theta_{213}\} $
$$A^{123} = \left(\begin{array}{cccccc}
1 & q_{23} & q_{23}q_{13} & q_{12}q_{13}q_{23} & q_{12}q_{13} & q_{12} \\
q_{32} & 1 & q_{13} & q_{13}q_{12} & q_{12}q_{13}q_{32} & q_{12}q_{32} \\
q_{32}q_{31} & q_{31} & 1 & q_{12} & q_{12}q_{32} & q_{12}q_{31}q_{32} \\
\cdot & \cdot & \cdot & 1 & q_{32} & q_{31}q_{32} \\
\cdot & \cdot & \cdot & q_{23} & 1 & q_{31} \\
\cdot & \cdot & \cdot & q_{13}q_{23} & q_{13} & 1
\end{array} \right) =
\left(\begin{array}{cc}
X & Y \\
\bar{Y} & \bar{X}
\end{array} \right) $$
where $\bar{X}^{T}=X$, $Y^{T}=Y$.
\rem{\bf Example 1.6.4.}\ \ Let $I=\{ 1,2,3\} $ and $\nu $ degenerate with
$\nu_{1}=2$, $\nu_{2}=0$, $\nu_{3}=1$. Then w.r.t. basis
$\{ \theta_{113},\theta_{131},\theta_{311}\} $
$$A^{113} = \left(\begin{array}{ccc}
1+q_{11} & q_{13}+q_{11}q_{13} & q_{13}^{2}+q_{11}q_{13}^{2} \\
q_{31}+q_{31}q_{11} & 1+q_{11}q_{13}q_{31} & q_{13}+q_{11}q_{13} \\
q_{31}^{2}+q_{31}^{2}q_{11} & q_{31}+q_{31}q_{11} & 1+q_{11} \\
\end{array} \right). $$
\subsection{A reduction to generic case } 

Some questions about the matrices $A^{(\nu )}$ for general $\nu $
(e.g. invertibility, positive definiteness) can be reduced to the
generic situation by using the following observation. Let $\nu
=\sum_{i}\nu_{i}i \in {\bf N}[I]$ be a degenerate weight.Let
$\tilde{I}$ be any set of size equal to $n=|\nu |=\sum_{i}\nu_{i}$
and let $\phi : \tilde{I}\longrightarrow I$ be a function which
maps exactly $\nu_{i}$ elements $ \tilde{i}$ of $\tilde{I}$ to $
i\in I$, and let ${\bf \tilde{q}}$ be the induced hermitian family
of parameters $ \tilde{q}_{\tilde{i},\tilde{j}}:= q_{i,j}
(\tilde{i},\tilde{j} \in \tilde{I})$ where $i=\phi ( \tilde{i}),
j=\phi (\tilde{j})$.Let ${\bf \tilde{f}}$ be the free associative
algebra with generators $\tilde {\theta }_{1},\dots ,\tilde{\theta
}_{n}$ and let $(\ ,\ )_{{\bf \tilde{q}}}$ be the sesquilinear
form on ${\bf \tilde{f}}$ associated to ${\bf \tilde{q}}$ (as in
1.3). Let ${\bf \tilde{f}}_{\tilde{\nu }}$ be the generic weight
space corresponding to $\tilde{\nu }\in {\bf N}[\tilde{I}]$ where
$\tilde{\nu }_{ \tilde{i}}=1$, for every $\tilde{i}\in \tilde{I}$.
Let $H=H_{\nu }$ be the group of all bijections of $\tilde{I}$
which map $\phi^{-1}\{ i\} $ to itself for every $i\in \phi(\tilde
I)$. This group is isomorphic to the Young subgroup
$\prod_{i}S_{\nu_{i}} \subset S_{n}$. Let Y be the subspace of
${\bf \tilde{f}}_{\tilde{\nu }}$ spanned by $H$-{\em invariant
vectors} $\tilde{\theta }_{H\tilde{{\bf i}}}=\sum_{h\in
H}\tilde{\theta }_{h\cdot \tilde{{\bf i}}}$ where $\tilde{\theta
}_{h\cdot \tilde{{\bf i}}} = \tilde\theta_{\tilde{ i}_
{h^{-1}(1)}}\cdots \tilde\theta_{\tilde{i}_{h^{-1}(n)}}$. Then for
the operator ${\bf \tilde{A}}$ associated to the form $(\ ,\
)_{\bf \tilde{q}}$ we have $${\bf \tilde{A}}(\tilde{\theta
}_{H\tilde{{\bf j}}})=\sum_{h\in H} {\bf \tilde{A}}(\tilde{\theta
}_{h\cdot {\bf \tilde{{\bf j}}}})=\sum_{h\in H}\sum_{{\bf
\tilde{i}}} (\tilde{\theta }_{h\cdot {\bf \tilde{{\bf
j}}}},\tilde{\theta }_{{\bf \tilde {{\bf i}}}})_{{\bf
\tilde{q}}}\tilde{\theta }_{{\bf \tilde{{\bf i}}}}.$$ By
Prop.1.6.1. we can write ${\bf \tilde{A}}(\tilde{\theta }_{H{\bf
\tilde{j}}})=\sum_{{\bf \tilde{i}}} A_{{\bf i,j}}^{(\nu
)}\tilde{\theta }_{{\bf \tilde{i}}}=\sum_{{\bf i}} A_{{\bf
i,j}}^{(\nu )}\tilde{\theta }_{H{\bf \tilde{i}}}$.Thus we have
proved that Y is an invariant subspace of the operator ${\bf
\tilde{A}}$ associated to the form $(\ ,\ )_{{\bf \tilde{q}}}$ and
moreover that the matrix of ${\bf \tilde{A}}|Y$ w.r.t basis of
$H$-invariant vectors $\tilde\theta_{H\bf \tilde i}$ coincides
with $A^{(\nu )}$. From this fact we conclude that

\begin{enumerate}
\item[1)] If ${\bf \tilde{A}}|_{{\bf \tilde{f}}_{\tilde{\nu }}}$ is
invertible, then ${\bf A}^{(\nu )}$ is invertible too. In particular
$[A^{(\nu )}]^{-1} _{{\bf i,j}}=\sum_{h\in H}[\tilde{A}^{(\tilde{\nu
})}]^{-1}_{{\bf \tilde{i}}, h{\bf \tilde{j}}}$,where ${\bf
\tilde{i},\tilde{j}}$ are chosen so that $\phi ({\bf \tilde{i}})={\bf
i}, \phi ({\bf \tilde{j}})={\bf j}$.This means that the entries of
$[A^{(\nu)}]^{-1}$($\nu$ degenerate)can be read off from the sums of
H-equivalent columns of the matrix $[\tilde{A}^{(\tilde{\nu
})}]^{-1}$($\tilde{\nu}$ generic). \item[2)] The determinant of $A^{(\nu
)}$ divides the determinant of $\tilde{A}^{(\tilde{\nu })}$.\item[3)] If
$\tilde{A}^{(\tilde{\nu })}$ is positive definite, then $A^{(\nu )}$ is
positive definite too.
\end{enumerate}

\subsection{Factorization of matrices $A^{(\nu )}$ for $\nu $ generic} 

                                First of all we point out that the rows of our
multiparametric matrices $A^{(\nu )}$ are not equal up to reordering
(what was true in [Zag], where all $q_{ij}$ are equal to q). Therefore,
the factorization of the matrices $A^{(\nu )}$ can not be reduced to the
factorization of the corresponding group algebra elements as was treated
by Zagier. Instead, by a somewhat tricky extension of the Zagier's
method  we show how this can be done on the matrix level. This is
achieved by studying a $q_{ij}$-deformation of the regular
representation  of the symmetric group which is only
quasimultiplicative, i.e., multiplicative only up to factors which are
diagonal ($q_{ij}$-dependent) matrices ("projective representation").
For $\nu =\sum \nu_{i}i \in {\bf N}[I]$ generic,$n=|\nu |=\sum \nu_{i}$
\ \  let $R_{\nu }$ denotes the action of the symmetric group $S_{n}$ on
the (generic) weight space ${\bf f}_{\nu }$, given on the basis $B_{\nu
}=\{ \theta_{{\bf i}}= \theta_{i_{1}}\cdots \theta_{i_{n}}, |{\bf
i}|=\nu \} $ of ${\bf f}_{\nu }$ by place permutations:$R_{\nu }(g):
\theta_{{\bf j}}=\theta_{j_{1}}\cdots \theta_{j_{n}} \longrightarrow
\theta_{g\cdot {\bf j}}=\theta_{j_{g^{-1}(1)}}\cdots
\theta_{j_{g^{-1}(n)}}.$Then $R_{\nu }$ is equivalent to the {\em right
regular representation} $R_{n}$ of $ S_{n}$.The corresponding {\em
matrix representation}, also denoted by $R_{\nu }(g)$, is given by
$R_{\nu }(g)_{{\bf i,j}}:=\delta_{{\bf i},g\cdot {\bf j}}.$ Now, we need
more notations.\\ Let $Q_{a,b}^{\nu }$ for $1\leq a, b\leq n$ and
$Q^{\nu }(g)$, for $g\in S_{n}$ be the following diagonal matrices
(multiplication operators on ${\bf f}_{\nu }$ ) defined by
$$
(Q_{a,b}^{\nu })_{{\bf i,i}}:=q_{i_{a}i_{b}};
\ {\rm e.g\ } (Q_{2,4}^{1234})_{4123,4123}=q_{13} {\rm\ if\ } I=\{ 1,2,3,4\},
\nu_{1}=\nu_{2}=\nu_{3}=\nu_{4} =1).
$$
$$
 Q^{\nu}(g)_{{\bf i,i}}:=q_{{\bf i},g^{-1}}=
        \prod_{a<b, g^{-1}(a)>g^{-1}(b)}q_{i_{a}i_{b}}\quad
(\Longrightarrow Q^{\nu}(g)=\prod_{(a,b)\in I(g^{-1})}Q^{\nu}_{a,b} ).
$$

Note that $\bar{q}_{ij}=q_{ji}$ imply that
$Q^{\nu}_{b,a}=[Q^{\nu}_{a,b}]^{*}$. We also denote by $|Q^{\nu}_{a,b}|$
the diagonal matrix defined by $|Q^{\nu}_{a,b}|_{{\bf
i,i}}=|q_{i_{a}i_{b}}|$. The quantity $Q^{\nu}_{a,b}\cdot
Q^{\nu}_{b,a}(=|Q^{\nu}_{a,b}|^{2})$ we abbreviate as $Q^{\nu}_{\{ a,b\}
}$.  More generally, for any subset $T\subseteq \{ 1,2,\cdots ,n\} $ we
shall use the notations

$$Q^{\nu}_{T}:=\prod_{a,b\in T, a\neq
b}Q^{\nu}_{a,b},\ \ \  \hbox{\large$\Box$}^{\nu}_{T} := I-Q^{\nu}_{T}
$$

\noindent (e.g. $Q^{\nu}_{\{ 3,5,6\} }=Q^{\nu}_{\{ 3,5\} }Q^{\nu}_{\{
3,6\} }Q^{\nu} _{\{ 5,6\}
}=Q^{\nu}_{3,5}Q^{\nu}_{5,3}Q^{\nu}_{3,6}Q^{\nu}_{5,6} Q^{\nu}_{6,5}$).
The following $q_{ij}$-{\em deformation of the representation} $R_{\nu
}$, defined by \ \ $\hat{R}_{\nu }(g):=Q^{\nu}(g)R_{\nu }(g), \ g\in
S_{n}$ \ will be crucial in our method for factoring the matrices
$A^{(\nu )}$ $\nu $-generic.

\bthm{PROPOSITION 1.8.1.} If $\nu $ is a generic weight with $|\nu |=n$, then
for the matrix $A^{(\nu )}$ of $(\ ,\ )_{\bf q}$ on ${\bf f}_\nu$ we have
$$A^{(\nu )}=\sum_{g\in S_{n}}\hat{R}_{\nu }(g)$$
\ethm
\vskip-1ex
{\bf Proof.}  The (${\bf i,j}$)-th entry of the r.h.s. is equal to
$\sum_{g\in S_{n}}\hat{R}_{\nu }(g)_{{\bf i,j}}=\\ \sum_{g\in
S_{n}}Q(g)_{{\bf i,i}}\hat{R}_{\nu }(g)_{{\bf i,j}}= \sum_{g\in
S_{n}}q_{{\bf i},g^{-1}}\delta_{{\bf i},g\cdot {\bf j}}= q_{{\bf i},\tau
^{-1}}$, if ${\bf i}=\tau {\bf j}$ (such $\tau $ is unique, because
$|{\bf i}|=|{\bf j}|=\nu $ is generic), what is just $A^{(\nu )} _{{\bf
i,j}}$, according to Prop.1.6.1 ii) and the proof follows.~\qed

Before we proceed with factorization of matrices $A^{(\nu )}$ we need
more detailed informations concerning our "projective" right regular
representation $\hat{R}_{\nu }$:

\vskip3pt\noindent
{PROPERTY 0.} (quasimultiplicativity) $\hat{R}_{\nu }(g_{1})\hat{R}_{\nu
}(g_{2})=\hat{R}_{\nu }(g_{1}g_{2}) \hbox{ \ if \ }
l(g_{1}g_{2})=l(g_{1})+l(g_{2})$,where $l(g):=Card\ I(g)$ is the lenght
of $g\in S_{n}$. This property follows from the following
general formula :

\bthm{PROPOSITION 1.8.2.}\hspace{-2.5mm}
For $g_{1},g_{2}\in S_{n}$ we have
$\hat{R}_{\nu }(g_{1})\hat{R}_{\nu }(g_{2})=M_{\nu }(g_{1},g_{2})
\hat{R}_{\nu }(g_{1}g_{2})$ where the multiplication factor is the
diagonal matrix $$M_{\nu }(g_{1},g_{2})=\prod_{(a,b)\in
I(g_{1}^{-1})-I(g_{2}^{-1}g_{1}^{-1})} Q^{\nu}_{\{ a,b\}
}\quad(=\prod_{(a,b)\in I(g_{1})\cap I(g_{2}^{-1})}Q^{\nu}_ {\{
g_{1}(a), g_{1}(b)\} }).$$
\ethm

For $1\leq a\leq b\leq n$ we denote by $t_{a,b}$ the following cyclic
permutation in $S_{n}$
$$t_{a,b}:=\left( \begin{array}{cccc}
a & a+1 & \cdots & b \\
b & a & \cdots & b-1
\end{array} \right)
$$
\noindent
which maps $b$ to $b-1$ to $b-2$ $\cdots $ to $a$ to $b$ and fixes all
$1\leq k <a$ and $b<k\leq n$.We also denote by $t_{a}:=t_{a,a+1} (1\leq
a<n)$ the transposition of adjacent letters $a$ and $a+1$.Then, from
Proposition 1.8.2, one gets the following more specific properties of
$\hat{R}_{\nu }$ which we shall need later on:

\vskip3pt\noindent
PROPERTY 1. (braid relations)
$$\hat{R}_{\nu }(t_{a})\hat{R}_{\nu }(t_{a+1})\hat{R}_{\nu
}(t_{a})= \hat{R}_{\nu }(t_{a+1})\hat{R}_{\nu }(t_{a})\hat{R}_{\nu
}(t_{a+1}), \ {\rm for}\ a=1,\dots ,n-2.
$$
$$
\hat{R}_{\nu }(t_{a})\hat{R}_{\nu }(t_{b})=\hat{R}_{\nu }(t_{b})
\hat{R}_{\nu }(t_{a}),
{\ \rm for\ } a,b \in  \{ 1,\dots ,n-1 \}  {\ \rm with\ } |a-b|\geq 2.
$$

\noindent
PROPERTY 2.$\hat{R}_{\nu }(g)\hat{R}_{\nu
}(t_{k,m})=\hat{R}_{\nu }(gt_{k,m}),$ for $g\in S_{m-1}\times S_{n-m+1},
1\leq k\leq m\leq n$.

\vskip3pt\noindent
PROPERTY 3. (commutation rules) i) For $1\leq a\leq a^{'}<m\leq n$
$$
\hat{R}_{\nu }(t_{a^{'},m})\hat{R}_{\nu }(t_{a,m})=Q^{\nu}_{\{ m-1,m\}
}\hat{R} _{\nu }(t_{a,m-1})\hat{R}_{\nu }(t_{a^{'}+1,m}).
$$
ii) Let
$w_{n}=n n-1\cdots 2 1$ be the longest permutation in $S_{n}$. Then for
any $g\in S_{n}$
$$
\hat{R}_{\nu}(gw_{n})\hat{R}_{\nu}(w_{n})=\hat{R}_{\nu}(w_{n})\hat{R}
_{\nu}(w_{n}g)=\Big(\prod_{a<b, g^{-1}(a)<g^{-1}(b)}Q^{\nu}_{\{
a,b\}}\Big)\hat{R}(g)
$$

\bthm{PROPOSITION 1.8.3.} For $ m\leq n$,  let $A^{(\nu
),m}:=\hat{R}_{\nu }(t_{1,m})+\hat{R}_{\nu }(t_{2,m})+\cdots +\hat{R}_
{\nu }(t_{m,m})\  (A^{(\nu ),1}=I)$. Then we have the following
factorization
$$
A^{(\nu )}=A^{(\nu ),1}A^{(\nu ),2}\cdots A^{(\nu ),n}.
$$
\ethm \vskip1pt
We now make a second reduction by expressing the matrices $A^{(\nu ),m}$ in
turn as a product of yet simpler matrices.
\bthm{PROPOSITION 1.8.4.}
Let $C^{(\nu ),m} (m\leq n)$ and $D^{(\nu ),m}
(m<n)$ be the following matrices \
$C^{(\nu ),m}:=[I-\hat{R}_{\nu }(t_{1,m})][I-\hat{R}_{\nu }(t_{2,m})]\cdots
[I-\hat{R}_{\nu }(t_{m-1,m})]$,\newline
$D^{(\nu ),m}:=[I-Q^{\nu}_{\{ m,m+1\} }\hat{R}_{\nu }(t_{1,m})]
[I-Q^{\nu}_{\{ m,m+1\} }\hat{R}_{\nu }(t_{2,m})]\cdots
[I-Q^{\nu}_{\{ m,m+1\} }\hat{R}_{\nu }(t_{m,m})]$.\ Then
$$A^{(\nu ),m}=D^{(\nu ),m-1}[C^{(\nu ),m}]^{-1}$$
\ethm
\subsection{Formula for the determinant of $A^{(\nu )}$, $\nu $ generic.} 

So far we have expressed the matrix $A^{(\nu )}$ as a product of
matrices like $[I-\hat{R}_{\nu}(t_{k,m})]^{-1}$ or
$I-Q^{\nu}_{\{ m,m+1\} }\hat{R}_{\nu}(t_{k,m})$ . Thus, in order to evaluate $\det
A^{(\nu )}$, we first compute the determinant of such matrices.

\bthm{LEMMA 1.9.1.} For $\nu $ generic with $|\nu |=n$, we have\newline
a) $\dsty\det(I-\hat{R}_{\nu }(t_{a,b}))=\prod_{\mu\subseteq \nu ,
|\mu|=b-a+1} (\Box_{\mu})^{(b-a)!(n+a-b-1)!} , ( a<b\leq n)$\newline b)
$\dsty\det(I-Q^{\nu}_{\{ b,b+1\} }\hat{R}_{\nu}(t_{a,b}))=
\prod_{\mu\subseteq \nu , |\mu|=b-a+2}(\Box_{\mu})
^{(b-a+2)(b-a)!(n+a-b-2)!}, (a\leq b<n)$\newline where for any subset
$T\subset I$ we denote by $\Box_{T}$ the quantity $$\Box_{T}:=1-{q}_{T}
; \qquad {q}_{T}\:=\prod_{i\neq j\in T}q_{ij}\qquad (=\prod_{\{ i\neq j\} \subset
T}|q_{ij}|^{2})$$ in which the last product is over all two-element
subsets of $T$ (We view $\nu $ as a subset of $I$, hence $\mu\subseteq
\nu $ means that $\mu$ is a subset of $\nu$). \ethm\vskip1pt


{\bf Proof.} a) Let $H:=<t_{a,b}>\subset S_{n}$ be the cyclic subgroup
of $S_{n}$ generated by the cycle $t_{a,b}$. Then, each $H$-orbit on
${\bf f}_{\nu }$, ${\bf f}_{\nu }^{[{\bf i}]_ {a}^{b}}=span\{
\theta_{t_{a,b}^{k}\cdot {\bf i}} | 0\leq k\leq b-a\} $, (which clearly
corresponds to a cyclic $t_{a,b}$-equivalence class $[{\bf i}]^{b}_{a}
=i_{1}\cdots (i_{a}i_{a+1}\cdots i_{b})\cdots i_{n}$ of the sequence
${\bf i}=i_{1}\dots i_{n}$ of weight $\nu $) is an invariant subspace of
$R_{\nu }(t_{a,b})$ (and hence of $\hat{R}_{\nu }(t_{a,b})$). Note that
$\hat{R} _{\nu }(t_{a,b})(\theta_{t_{a,b}^{k}\cdot {\bf
i}})=c_{k}\theta_{t_{a,b}^{k+1} \cdot {\bf i}}$ where
$c_{k}=q_{t_{a,b}^{k}\cdot {\bf i}, t_{a,b}^{-1}} (0\leq k\leq b-a)$ i.e.

$$\left\{\begin{array}{lcl}
c_{0}&=&q_{i_{a}i_{b}}q_{i_{a+1}i_{b}}\cdots q_{i_{b-1}i_{b}},\\
c_{1}&=&q_{i_{a+1}i_{a}}q_{i_{a+2}i_{a}}\cdots q_{i_{b}i_{a}},\\
&\vdots&\\
c_{b-a}&=&q_{i_{b}i_{b-1}}q_{i_{a}i_{b-1}}\cdots q_{i_{b-2}i_{b-1}}.
\end{array}\right.
$$


Thus
\ $\hat{R}_{\nu }(t_{a,b})|f_{\nu }^{[{\bf i}]_{a}^{b}}$ is a cyclic
operator, and

\begin{eqnarray*} \det(I-\hat{R}_{\nu }(t_{a,b})|f_{\nu }^{[{\bf
i}]_{a}^{b}})&=&1-c_{0}c_{1} \cdots c_{b-a}=1-\prod_{i\neq j\in \{
i_{a},\dots ,i_{b}\} }q_{ij}= \Box_{\{ i_{a},\dots ,i_{b}\} }.
\end{eqnarray*}
Note that this determinant depends only on the set $ \{ i_{a},i_{a+1},
\dots, i_{b}\} $ and that there are $(b-a)!(n-(b-a+1))!$ cyclic
$t_{a,b}$ -equivalence classes corresponding to any given
$(b-a+1)$-set $\mu=\{ i_{a},\dots ,i_{b}\} \subset \nu $.
(Here we identify a generic weight $\nu =\sum \nu_{i}\cdot
i$, $\nu_{i}\leq 1$ with the set $\{ i\in I|\nu_{i}=1\} $).
b) Quite analogous as a).~\qed

\bthm{THEOREM 1.9.2.[DETERMINANT FORMULA]}.
For $\nu$  generic, we have
$$\det A^{(\nu )}=\prod_{\mu\subseteq \nu, |\mu| \geq 2
}(\Box_{\mu})^{(|\mu|-2)!(|\nu |-|\mu|+1)!}.$$ \ethm
\noindent
In particular, in Example 1.6.3 we have $$\det
A^{123}=(1-|q_{12}|^{2})^{2}(1-|q_{13}|^{2})^{2}
(1-|q_{23}|^{2})^{2}(1-|q_{12}|^{2}|q_{13}|^{2}|q_{23}|^{2})$$ {\bf
Remark 1.9.3.}Theorem 1.9.2 is a multiparametric extension of Theorem 2
in [Zag] which (case $(q_{ij}=q)$) reads as:$\det
A_{n}(q)=\prod_{k=2}^{n}(1-q^{k(k-1)})^{\frac{n!(n-k+1)}{k(k-1)}}$(e.g.
$\det A_{3}(q)=(1-q^{2})^{6}(1-q^{6})$).\vskip1pt

\bthm{THEOREM 1.9.4.}The matrix $A=A({\bf q})$ associated to the sesquilinear
form $(\ ,\ )_{{\bf q}}$ on ${\bf f}$,(see 1.3 and 1.6) is
positive definite if $|q_{ij}|<1$,for all $i,j\in I$,so that the
$q_{ij}$-cannonical commutation relations 1.1(1) have a Hilbert space
realization (cf.1.5).
\ethm

\section{Formulas for the inverse of $A^{(\nu )}$, $\nu $ generic.} 

 The problem of computing the inverse of matrices $A^{(\nu )}$
appears in the expansions of the number operators and transition
operators (c.f [MSP]). It is also related to a random walk problem
on symmetric groups and in several other situations (hyperplane
arrangements, contravariant forms on certain quantum groups).
We shall give here two types of formulas for $[A^{(\nu )}]^{-1}$: Zagier
type formula and Bo\v zejko-Speicher type formulas.

\subsection{Zagier type formula} 

First we give a formula for the inverse of $A^{(\nu )}$, $\nu $ generic,
which follows from Prop.1.8.3 and Prop.1.8.4 :
\begin{eqnarray*}
[A^{(\nu )}]^{-1}&=&[A^{(\nu ),n}]^{-1}\cdots [A^{(\nu ),1}]^{-1}\\
&=&C^{(\nu ),n}\cdot [D^{(\nu ),n-1}]^{-1}\cdot C^{(\nu ),n-1}\cdot [D^{(\nu
),n-2}]^{-1}\cdots C^{(\nu ),2}\cdot [D^{(\nu ),1}]^{-1}
\end{eqnarray*}
To invert $A^{(\nu )}$, therefore, the first step is to invert $D^{(\nu )
,m}$ for each $m<n$.
Then one can use multiparametric
extensions of Propositions 3. and 4. of [Zag] which are to long to state
them here.

\subsection{Bo\v zejko-Speicher type formulas} 

In addition to the, multiplicative in spirit, Zagier type formula for
the inverse of $A^{(\nu )}$ ($\nu $ generic), given in 2.1., one also
has another, additive in spirit, Bo\v zejko-Speicher type formula
(c.f. [BSp1], Lemma 2.6.) which, in the case of symmetric group
$S_{n}$, we shall present here, in slightly different notation,
together with several improvements.
For $J=\{ j_{1}<j_{2}<\cdots <j_{l-1}\} \subseteq
\{ 1,2,\dots ,n-1\} $ let
$S_{J}$ be the following Young subgroup of $S_{n}$ defined by
$S_{J}:=S_{j_{1}}\times S_{j_{2}-j_{1}}\times \cdots \times S_{n-j_{l-1}},\
\ S_{\phi }=S_{n}.$
Then the following is the left coset decomposition:
$S_{n}=\gamma_{J}S_{J}$,
where $\gamma_{J}=\{ g\in S_{n}| g(1)<g(2)<\cdots <g(j_{1}), g(j_{1}+1)<
\cdots <g(j_{2}),\cdots ,g(j_{l-1}+1)<\cdots <g(n)\} $.
The definition of $\gamma_{J}$ can also be put in the way

\rem{\bf FACT 2.2.1.}
$g\in \gamma_{J}\Leftrightarrow g(1)g(2)\cdots g(n)$ is the
shuffle of the sets \newline
$[1..j_{1}], [j_{1}+1..j_{2}],\dots [j_{l-1}+1,n]\Leftrightarrow $ the
descent set $Des(g)=\{ 1\leq i\leq n-1 | g(i)>g(i+1)\} $ of $g$ is contained
in the set $J$ (c.f. [Sta, pp. 69-70]). (Here $[a..b]$ denotes the set
$\{ a,a+1,\dots ,b\} $.)
Moreover, each $g\in S_{n}$ has the unique factorization $g=a_{J}g_{J}$
with $g_{J}\in S_{J}$ and $a_{J}\in \gamma_{J}$ and with $l(g)=l(a_{J})
+l(g_{J})$.
For arbitrary subset $X\subseteq S_{n}$ we define the matrix $\hat{R}_{\nu }
(X)$ by

$$\hat{R}_{\nu }(X):=\sum_{g\in X}\hat{R}_{\nu }(g)$$
\bthm{PROPOSITION 2.2.2.} Let $\nu $ be a generic weight, $|\nu |=n$. For any
subset
$J=\{ j_{1}<j_{2}<\cdots j_{l-1}\} $ of $\{ 1,2,\dots ,n-1\} $
let $A^{(\nu )}_{J}, \Gamma
^{(\nu )}_{J}$ be the following matrices

$$A^{(\nu )}_{J}=\hat{R}_{\nu }(S_{J}) (=\sum_{g\in S_{J}}\hat{R}_{\nu
}(g)),\quad \Gamma^{(\nu )}_{J}=\hat{R}_{\nu }(\gamma_{J}) (=\sum_{g\in
\gamma_{J}} \hat{R}_{\nu }(g)).$$
\noindent
Then the matrix $A^{(\nu )} (=A^{(\nu )}_{\phi })$ of the sesquilinear
form $(\ ,\ )_{{\bf q}}$ (see Prop.1.8.1) has the following
factorizations $$\belowdisplayskip0pt A^{(\nu )}=\Gamma^{(\nu
)}_{J}A^{(\nu )}_{J}\quad (\Rightarrow \Gamma^{(\nu )} _{J}=A^{(\nu
)}[A^{(\nu )}_{J}]^{-1})$$ \ethm

{\bf Proof.} By quasimultiplicativity of $\hat{R}_{\nu }$ and FACT 2.2.1.~\qed

The following formula is the Bo\v zejko-Speicher adaptation of an Euler-type
character formula of Solomon. In the case $W=S_{n}$ it reads as follows :

\bthm{LEMMA 2.2.3.} (c.f. [BSp2] Lemma 2.6) Let $w_{n}=n n-1\dots 1$ be the
longest permutation in $S_{n}$. Then we have
$$\belowdisplayskip0pt \sum_{J\subseteq \{ 1,2,\dots ,n-1\}
}(-1)^{n-1-|J|}\Gamma^{(\nu )}_{J}= \hat{R}_{\nu }(w_{n})$$ \ethm

 For reader's convenience we include here a variant of the proof
(our notation is slightly different). For any subset $M\subseteq \{ 1,2,\dots
,n-1\} $ we denote by $\delta_{M}$  the subset of $S_{n}$ consisting of all
permutations  $g\in S_{n}$ whose descent set $Des(g)$ is equal to M. Then by
FACT 2.2.1  it is clear that $\gamma_{J}=
\bigcup_{M\subseteq J}\delta_{M}$ (disjoint union), implying that

$$\hat{R}_{\nu }(\gamma_{J})=\sum_{M\subseteq J}\hat{R}_{\nu }(\delta_{M})$$
By the inclusion-exclusion principle we obtain

$$\hat{R}_{\nu }(\delta_{M})=\sum_{J\subseteq M}(-1)^{|M-J|}\hat{R}_{\nu }
(\gamma_{J})$$
By letting $M=\{ 1,2,\dots ,n-1\} (\Rightarrow \delta_{M}=\{ w_{n}\} )$ we
obtain the desired identity.
By combining Prop.2.2.2. and Lemma 2.2.3 we obtain the following relation
among the inverses of matrices $A^{(\nu )}_{J}$'s.

\bthm{PROPOSITION 2.2.4. (Long recursion for the inverse of $A^{(\nu )}$):}
We have
$$[A^{(\nu )}]^{-1}=(\sum_{\emptyset \neq J\subseteq \{ 1,2,\dots ,n-1\} }
(-1)^{|J|+1}[A^{(\nu )}
_{J}]^{-1})(I+(-1)^{n}\hat{R}_{\nu }(w_{n}))^{-1}$$
\ethm
\rem{REMARK 2.2.5.}  Let us associate to each subset $\phi \neq J=\{
j_{1}<j_{2}< \cdots <j_{l-1}\} \subseteq \{ 1,2,\dots ,n-1\} $ a {\em
subdivision} $\sigma (J)$ of the set $\{ 1,2,\dots ,n\} $ into intervals
by $\sigma (J)= J_{1}J_{2}\cdots J_{l}$, where
$J_{k}=[j_{k-1}+1..j_{k}] (j_{0}=1,j_{l}=n)$. (Here $[a..b]$ denotes the
interval $\{ a,a+1,\dots ,b-1,b\} $ and abbreviate $[a..a] (=\{ a\} )$
to $[a]$). The Young subgroup $S_{J}$ can be written as direct product
of commuting subgroups $$S_{J}=S_{[1..j_{1}]}S_{[j_{1}+1..j_{2}]}\cdots
S_{[j_{l-1}+1..n]} =S_{J_{1}}S_{J_{2}}\cdots S_{J_{l}}$$ where for each
interval $I=[a..b]$, $1\leq a\leq b\leq n$ we denote by
$S_{I}=S_{[a..b]}$ the subgroup of $S_{n}$ consisting of permutations
which are identity on the complement of $[a.. b]$ (i.e.
$S_{[a..b]}=S_{1}^{a-1}\times S_{b-a+1}\times S_{1}^{n-b}$). By denoting
accordingly $A^{(\nu )}_{I}=A^{(\nu )}_{[a..b]}:= \hat{R}_ {\nu
}(S_{[a..b]})$, we can rewrite the formula for $[A^{(\nu
)}]^{-1}=[A^{(\nu )}_{[1..n]}]^{-1}$ in Prop.2.2.4. as follows:

$$[A^{(\nu )}_{[1..n]}]^{-1}=(\sum_{\sigma =J_{1}\cdots
J_{l},l\geq 2}(-1)^{l}[A^{(\nu )}_{J_{1}}]^{-1}\cdots [A^{(\nu
)}_{J_{l}}]^{-1})(I+(-1)^{n}\hat{R}_{\nu }(w_{n}))^{-1}\eqno(*)$$
where the sum is over all subdivisions of the set $\{ 1,2,\dots ,n\} $.
Similar formula we can write for $[A^{(\nu )}_{[a..b]}]^{-1}$ for any
nondegenerate interval $[a..b]$, $1\leq a<b\leq n$. Of course if $a=b,
[A^{(\nu )}_{[a..b]}]^{-1}$ is the identity matrix.
Now we shall use an ordering denoted by $\prec$ on the set $\Sigma_{n}$ of
all subdivisions of the set $\{ 1,2,\dots ,n\} $, called {\em reverse
refinement order}, defined by $\sigma \prec \sigma ^{'}$ if $\sigma ^{'}$ is
finer than $\sigma $ i.e. $\sigma ^{'}$ is obtained by subdividing each
nontrivial interval in $\sigma $. The minimal and maximal elements in
$\Sigma_{n}$ are denoted by $\hat 0_{n}(=[1..n])$ and ${\hat
1_{n}}=[1][2]\cdots [n]$. We shall call $(\Sigma_n,\prec)$ the {\em lattice
of subdivisions} of $\{1,2,\dots, n\}$. For example we have
$\Sigma_{1}=\{ [1]\}$, $\Sigma_{2}=\{ [12], [1][2]\}$, $\Sigma_{3}=\{
[123], [1][23], [12][3], [1][2][3]\}$, $\Sigma_{4}=\{ [1234],$ $[123][4],$
$[12][34]$, $[1][234]$, $[12][3][4]$, $[1][23][4]$, $[1][2][34]$, $[1][2][3][4]\}$
 (see Figure 1).\\
(Here $[1234]$ denotes the interval $[1..4]=\{ 1,2,3,4\} $ etc.)


\begin{figure}
\centerline{\includegraphics[width=12cm,height=8.307cm,keepaspectratio]{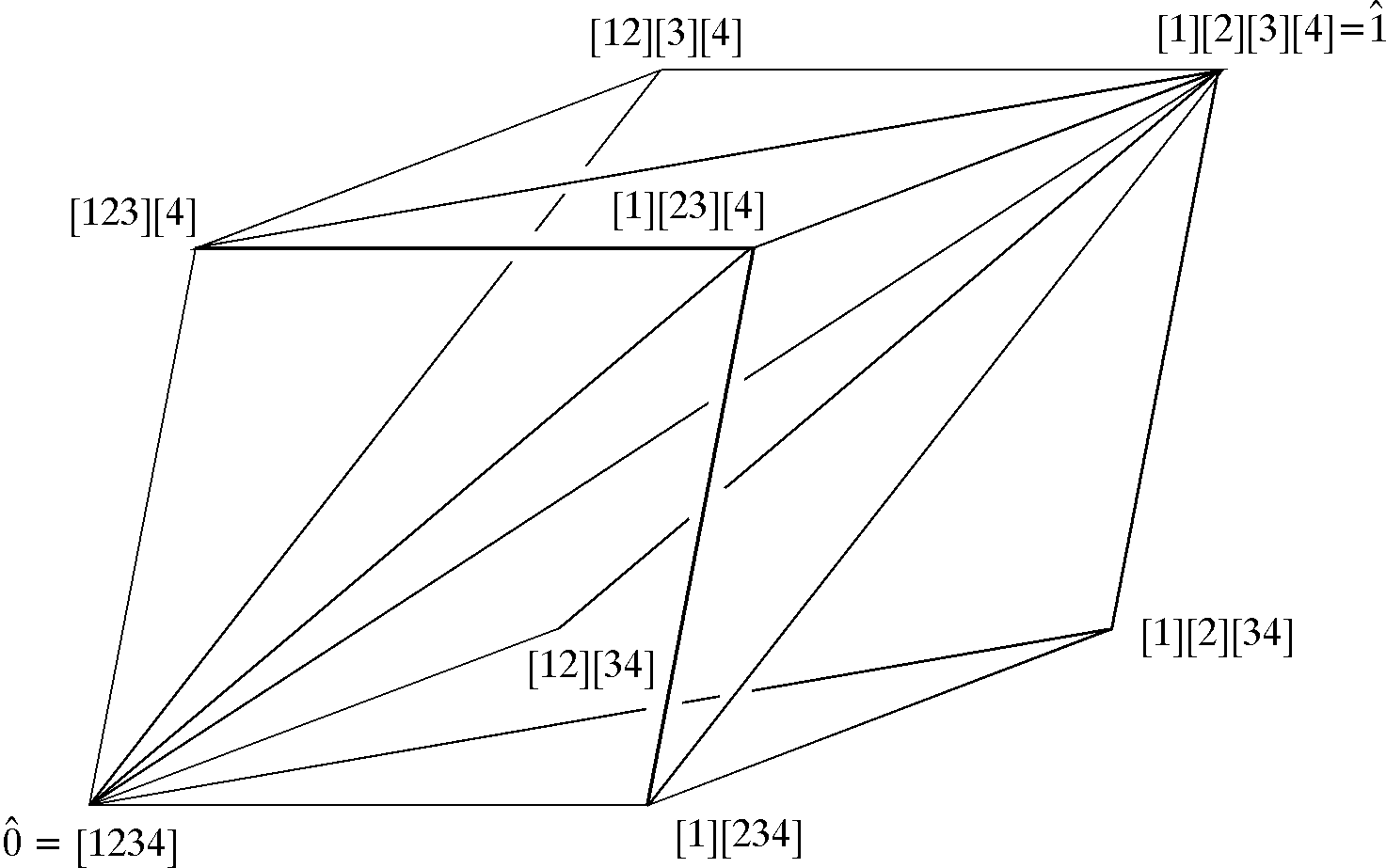}}
\caption
{$\Sigma_{4}=$ The lattice of subdivisions of $\{ 1,2,3,4\} $.}
\end{figure}

Now for each interval $I=[a..b], 1\leq a<b\leq n$ we
denote by $w_{I}=w_{[a..b]}:=1\ 2\cdots a-1\ b\ b-1\cdots a\ b+1\cdots n$
the longest permutation in $S_{[a..b]} (=S_{1}^{a-1}\times S_{b-a+1}
\times S_{1}^{n-b})$ and by $\Psi_{I}^{\nu}=\Psi^{\nu}_{[a..b]}, a<b$
the following matrix
\begin{eqnarray*} [I+(-1)^{b-a+1}\hat{R}_{\nu
}(w_{[a..b]})]^{-1}=\frac{1}{\bbox^{\nu}_{[a..b]}}[I-(-1)^{b-a+1}\hat{R}_{\nu
}(w_{[a..b]})]\\ =\frac{1}{\bbox_{I}^{\nu}}\Phi_{I}^{\nu},\quad {\rm
where} \qquad \ \Phi_{I}^{\nu}:= I-(-1)^{|I|}\hat{R}_{\nu}(w_{I})
\end{eqnarray*} and  where $\bbox^{\nu}_{[a..b]}$ is the diagonal matrix
(cf. the definition of $\bbox^{\nu}_{T}$ given in 1.8):

$$\bbox^{\nu}_{[a..b]}=\bbox^{\nu}_{\{ a,a+1,\cdots ,b\} }
=I-Q^{\nu}_{\{ a,a+1,\dots ,b\} }=I-\prod_{a\leq k<l\leq b}
|Q^{\nu}_{k,l}|^{2}, \ [Q^{\nu}_{k,l}]_{i_{1}\cdots i_{n},i_{1}\cdots i_{n}}=
q_{i_{k}i_{l}}.$$
Accordingly, for any subdivision $\sigma =I_{1}I_{2}\cdots I_{l} \in
\Sigma_{n}$ we define $\Psi^{\nu}_{\sigma }:=\prod_{j: |I_{j}|\geq
2}\Psi^{\nu}_{I_{j}}$ (factors commute here,because $I_{j}'$s are
disjoint!), and similarly for any chain ${\cal C}: \sigma^{(1)} \prec\cdots
\prec\sigma^{(m)}$ in $\Sigma_{n}$ we define
$$\Psi^{\nu}_{\cal C}\:= \overleftarrow\prod_{1\leq j\leq m}\Psi^{\nu}_
{\sigma^{(j)}} =
\Psi^{\nu}_{\sigma^{(m)}}\cdots \Psi^{\nu}_ {\sigma^{(1)}}
$$
In the same
way we introduce notations $\bbox^{\nu}_{\cal C}$ and $\Phi^{\nu}_ {\cal
C}$ and observe that then $\Psi^{\nu}_{\cal
C}=\frac{1}{\bbox^{\nu}_{\cal C}}\Phi^{\nu}_{\cal C}$. For example, if
${\cal C}: {\hat 0}_{5}=[12345] \prec [12][345] \prec [1][2][34][5] \prec {\hat 1}_{5}$,
then {\rm for \ any \ generic \  weight}\   $\nu , |\nu |=5$ we have
$\Psi^{\nu}_{\cal C}=\Psi^{\nu}_{\{ 3,4\} }(\Psi^{\nu}_{\{ 1,2\}
}\Psi^{\nu} _{\{ 3,4,5\} })\Psi^{\nu}_{\{ 1,2,3,4,5\} }\\
=\frac{1}{\bbox^{\nu}_{\{ 3,4\} }\bbox^{\nu}_{\{ 1,2\} } \bbox^{\nu}_{\{
3,4,5\} }\bbox^{\nu}_{\{ 1,2,3,4,5\} }}\Phi^{\nu}_{\{ 3,4\} }
\Phi^{\nu}_{\{ 1,2\} }\Phi^{\nu}_{\{ 3,4,5\} }\Phi^{\nu}_{\{ 1,2,3,4,5\}
}, $.

Now we can state our first explicit formula for the inverse of $A^{(\nu
)}$ in terms of the involutions $w_{I}=w_{[a..b]}, 1\leq a<b\leq n$.

\bthm{THEOREM 2.2.6.(INVERSION FORMULA -- CHAIN VERSION).}
 Let $\nu $ be a generic weight, $|\nu |=n$. Then

$$[A^{(\nu )}]^{-1}=\sum_{\cal C }(-1)^{b_{+}({\cal
C})+n-1}\Psi^{\nu}_{\cal C} =\sum_{\cal C}\frac{(-1)^{b_{+}({\cal
C})+n-1}}{\bbox^{\nu}_{\cal C}}\Phi^{\nu}_{\cal C}$$ where the summation
is over all chains ${\cal C}: {\hat 0_{n}}=
\sigma^{(0)} \prec \sigma^{(1)}\cdots \prec \sigma ^{(m)} \prec {\hat 1_{n}}$ in the
subdivision lattice $\Sigma_{n}$ and where $b_{+}({\cal C})$ denotes the
total number of nondegenerate intervals appearing in members of $\cal C
$. \ethm
{\bf Proof.}  The formula follows by iterating the formula (*)
in Remark 2.2.5.~\qed

\rem{REMARK 2.2.7.}  If we represent chains ${\cal C} :
{\hat 0}_{n}=\sigma^{(0)} \prec \sigma^{(1)} \prec \cdots \prec \sigma^{(m-1)} \prec {\hat
1}_{n}$ of length $m\geq 1$ as {\em generalized bracketing} (of {\em
depth} $m$) of the word $12\cdots n$ with one pair of brackets for each
nondegenerate interval appearing in the members of $\cal C $ (e.g.
${\hat 0}_{5} =[12345] \prec [12][345] \prec [1][2][34][5] \prec {\hat 1}_{5}$ is
represented as $[[12][[34]5]]$), then we can write the bracketing version
of the Inversion formula of Thm.2.2.6 as

$$[A^{(\nu )}]^{-1}=\sum_{\beta }(-1)^{b(\beta )+n-1}\Psi^{\nu}_{\beta }
=\sum_{\beta}\frac{(-1)^{b(\beta)+n-1}}{\bbox^{\nu}_{\beta}}\Phi^{\nu}_{\beta}
$$
where the sum is over all generalized bracketings of the word $12\cdots
n$ and where $b(\beta )$ denotes the number of pairs of brackets in
$\beta $ and where $\Psi^{\nu}_{\beta }:=\Psi^{\nu}_{\cal C }$,
$\Phi^{\nu}_{\beta} :=\Phi^{\nu}_{\cal C}$,
$\bbox^{\nu}_{\beta}:=\bbox^{\nu}_{\cal C}$ if $\beta $ is associated to
the (unique!) chain $\cal C$ in $\Sigma_{n}$ (e.g.
$\Psi^{\nu}_{[[12][[34]5]]}= \Psi^{\nu}_{[3..4]}
(\Psi^{\nu}_{[1..2]}\Psi^{\nu}_{[3..5]})\Psi^{\nu}_{[1..5]}=\Psi^{\nu}_{[1..2]}
\Psi^{\nu}_{[3..4]}\Psi^{\nu}_{[3..5]} \Psi^{\nu}_{[1..5]}$\newline
$\dsty=\frac{1}{\bbox^{\nu}_{\{ 1,2\} }\bbox^{\nu}_{\{ 3,4\} }
\bbox^{\nu}_{\{ 3,4,5\} } \bbox^{\nu}_{\{ 1,...,5\} }}(I-\hat{R}_{\nu
}(w_{[1..2]}))(I-\hat{R}_{\nu } (w_{[3..4]}))(I+\hat{R}_{\nu
}(w_{[3..5]}))(I+\hat{R}_{\nu }(w_{[1..5]}))$ .

In particular for Example 1.6.3 ($I=\{ 1,2,3\} ,
\nu_{1}=\nu_{2}=\nu_{3}=1$) we have
\noindent
\[\begin{array}{@{}l}
[A^{123}]^{-1} = -\Psi_{[123]}+\Psi_{[[12]3]}+\Psi_{[1[23]]}=
\frac{-1}{\bbox_{\{1,2,3\}}}(I-\hat{R}_{123}(321)) +
\frac{1}{\bbox_{\{1,2\}}\bbox_{\{1,2,3\}}}\\
(I+\hat{R}_{123}(213))(I-\hat{R}_{123}(321))+ \frac{1}{\bbox_{\{2,3\}}
\bbox_{\{1,2,3\}}}(I+\hat{R}_{123}(132))(I-\hat{R}_{123}(321)).
\end{array}
\]
Similarly for $I=\{ 1,2,3,4\}, \nu_{1}=\nu_{2}=\nu_{3}=\nu_{4}=1$ we have
\begin{eqnarray*}
[A^{1234}]^{-1}&=&\Psi_{[1234]}-\Psi_{[1[234]]}-\Psi_{[12[34]]}-
\Psi_{[1[23]4]}-\Psi_{[[12]34]}-\Psi_{[[123]4]}+\\
&+&\Psi_{[[12][34]]}+\Psi_{[[[12]3]4]}+\Psi_{[[1[23]]4]}+
\Psi_{[1[[23]4]]}+\Psi_{[1[2[34]]]}
\end{eqnarray*}
(Here we suppresed the upper indices in
$\Psi^{123}_{\beta}$ and $\Psi^{1234}_{\beta}$).

\bthm{COROLLARY 2.2.8. (EXTENDED ZAGIER'S CONJECTURE):} For $\nu $ generic,
$|\nu |=n$, for the inverse of the matrix $A^{(\nu)}=A^{(\nu)}({\bf q})$ we
have
$$[A^{(\nu)}]^{-1}\in \frac{1}{\bbox^{\nu}}{\rm Mat}_{n!}(Z[q_{ij}])\leqno i)$$
with $\bbox^{\nu}$ denoting the diagonal matrix
$\prod_{1\leq a<b\leq n}\bbox^{\nu}_{[a..b]}
=\prod_{1\leq a<b\leq n}(I-\prod_{a\leq k\neq l\leq b}Q^{\nu}_{k,l})$
$$[A^{(\nu)}]^{-1}\in \frac{1}{d_{\nu}}{\rm Mat}_{n!}(Z[q_{ij}])\leqno i')$$
where $d_{\nu}$ is the following quantity $\prod_{\mu\subseteq \nu
,|\mu|\geq 2}\Box_{\mu} =\prod_{\mu\subseteq \nu ,|\mu|\geq 2}
(1-\prod_{i\neq j\in \mu} q_{ij})$ \linebreak
($\Box_{\mu}$ and ${q}_{\mu}$ are
the same as in Lemma 1.9.1). In particular when all $q_{ij}=q$ (Zagier's
case) we have from i):
$$[A^{\nu}(q)]^{-1}\in \frac{1}{\delta_{n}(q)}{\rm Mat}_{n!}(Z[q]) \leqno ii)$$
where
$\delta_{n}(q)=\prod_{1\leq a<b\leq n}(1-q^{(b-a+1)(b-a)})
=\prod_{k=2}^{n}(1-q^{k(k-1)})^{n-k+1}$.
\ethm

{\bf Proof.} i) follows from Thm 2.2.6 by taking the common denominator
which turns out to be $\bbox^{\nu}=\prod_{1\leq a<b\leq n}\bbox^{\nu}_
{[a..b]}$ because any $\bbox^{\nu}_{[a..b]}$ appears at most once in each
of the denominators $\bbox^{\nu}_{\cal C}$ (and actually appears in at least
one of them).\newline
i') The entries of $\bbox^{\nu}$ are zero or $\bbox^{\nu}_{{\bf i,i}}$
where ${\bf i}=i_{1}\cdots i_{n}$ is any permutation of $\nu $
$(|{\bf i}|=\nu )$ considered as a subset of $I$ (because $\nu $ is
generic!). Since
\begin{eqnarray*}
\bbox^{\nu }_{{\bf i,i}}&=&\prod_{1\leq a<b\leq n}(1-\prod_{a\leq k\neq
l\leq b} q_{i_{k}i_{l}})=\prod_{1\leq a<b\leq n}\Box_{\{
i_{a},i_{a+1},\dots ,i_{b}\} } \end{eqnarray*} we see that
$\bbox^{\nu}_{{\bf i,i}}$ divides $d_{\nu}$.\newline ii) Note that in
case all $q_{ij}=q$:
\begin{eqnarray*}
\bbox^{\nu}_{{\bf i,i}}&=&\prod_{1\leq a<b\leq n}(1-\prod_{a\leq k\neq l\leq b}
q)=\prod_{k=2}^{n}(1-q^{k(k-1)})^{n-k+1}=\delta_{n}(q).
\end{eqnarray*}This completes the proof of Extended Zagier's conjecture.~\qed

\rem{REMARK 2.2.9.} In [Zag] p.201 Zagier conjectured that $A_{n}(q)^{-1}
\in \frac{1}{\triangle_{n}}{\rm Mat}_{n!}(Z[q])$, where $\triangle_{n}:=
\prod_{k=2}^{n}(1-q^{k(k-1)})$ and checked this conjecture for $n\leq 5$.
But we found that this conjecture failed for $n=8$
(see Examples to Prop.2.2.15). It seems that our
statement in Corollary 2.2.8 ii) is the right form of a conjecture valid for
all $n$ when all $q_{ij}$ are equal.
\bthm{PROPOSITION 2.2.10.} Let $c_{n}$ be the number ${\hat 0}_{n}-{\hat
1}_{n}$ chains in the subdivision lattice
$\Sigma_{n}$ (i.e. the number of $\Psi $-terms in the formula for $[A^{(\nu )}]
^{-1}$ $\nu $ generic, $|\nu |=n$ in Thm.2.2.6 ), $c_{0}:=0, c_{1}:=1$. Then
$$C(t)=\sum_{n\geq 0}c_{n}t^{n}=\frac{1}{4}(1+t-\sqrt
{1-6t+t^{2}})=t+t^{2}+3t^{3}+11t^{4}+45t^{5}+197t^{6}+\cdots
$$
\ethm
{\bf Proof.}  By Remark 2.2.7 this counting is equivalent to the Generalized
bracketing problem of Schr\"oder (1870) (see [Com], p.56). In fact, the
numbers $c_{n}$ can be computed faster via linear reccurence relation
(following from the fact that $C(t)$ is algebraic )
:$(n+1)c_{n+1}=3(2n-1)c_{n}-(n-2)c_{n-1}, \ \  n\geq 2, \ \
c_{1}=c_{2}=1$.~\qed
More generally,by a formal language method we found in [MS] that the number
$c_{n,k}$ of chains,as above, having alltogether $k$ nondegenerate intervals is
equal to  $c_{n,k}={{n+k+1} \choose {k}}{{n-2} \choose {k-1}}/{\it n}$\,\,(e.g.$ c_{3,1} =1,
c_{3,2}=2,c_{4,1}=1,c_{4,2}=5,c_{4,3}=5$ ) yielding a simple formula
for $ c_{n}=c_{n,1}+\cdots +c_{n,n-1}.$(This formula is relevant to non-intersecting
diagonals structures in one proof of Four Color Theorem and it is much simpler
than the one given in [CM1,CM2]).\\
Now we turn our attention to computation of entries in
the inverse of $A^{(\nu )}$, $\nu $ generic. First we note that any
$n!\times n!$ matrix $A$ can be written as $A=\sum_{g\in
S_{n}}A(g)R_{n}(g)$, where $A(g)$ are diagonal matrices defined by
$A(g)_{{\bf i,i}}=A_{{\bf i},g^{-1}\cdot {\bf i}}$ (all ${\bf i}$)
($R_{n}(g)$ is the right regular representation matrix $R_{n}(g)_{{\bf
i,j}}=\delta_{{\bf i},g\cdot {\bf j}}$, c.f. 1.8).We call $A(g)$ the
{\em g-th diagonal} of $A$. Hence, if we write
\begin{eqnarray*} A^{(\nu )}=\sum_{g\in S_{n}}A^{(\nu )}(g)R_{\nu
}(g),\quad [A^{(\nu )}]^{-1}=\sum_{g\in S_{n}}[A^{(\nu )}]^{-1}(g)R_{\nu
}(g) \end{eqnarray*} then by Prop.1.8.1 ($\nu $ generic ) we have
$$A^{(\nu )}(g)=Q^{\nu}(g)=\prod_{(a,b)\in I(g^{-1})}Q^{\nu}_{a,b}\  ; \
(Q^{\nu}_{a,b})_{{\bf i,i}}=q_{i_{a}i_{b}}$$ To compute $[A^{(\nu
)}]^{-1}(g)$ we first write
$[A^{(\nu )}]^{-1}(g)=\Lambda^{\nu }(g)A^{(\nu )}(g)$
where $\Lambda^{\nu }(g)$ are yet unknown diagonal matrices.

Similarly, for each $\emptyset \neq J\subseteq \{ 1,2,\dots ,n-1\} $
we write $[A^{(\nu )}_{J}]^{-1}(g)=\Lambda^{\nu}_{J}(g)A^{(\nu)}_{J}(g)$
 and for any $I=[a..b]\subseteq \{ 1,2,\dots ,n\} $ we write
$[A^{(\nu)}_{I}]^{-1}(g)=\Lambda^{\nu}_{I}(g)A^{(\nu)}_{I}(g)$
where $\Lambda^{\nu}_{J}(g)$ and $\Lambda^{\nu}_{I}(g)$ are unknown
diagonal matrices.

If $\sigma (J)=J_{1}J_{2}\cdots J_{l}$ is the subdivision
of $\{ 1,2,\dots ,n\} $ (cf. Remark 2.2.5) associated to $J$,
and if $g=g_{1}g_{2}\cdots g_{l} \in S_{J}=S_{J_{1}}S_{J_{2}}\cdots
S_{J_{l}}$, then $\Lambda^{\nu}_{J}(g)=\Lambda^{\nu}_{J_{1}}(g_{1})
\cdots \Lambda^{\nu}_{J_{l}}(g_{l})$.

Let us denote by $S^{>}_{n}$ (resp. $S^{<}_{n}$) the subset
of $S_{n}$ of all elements $g$ such that $g(1)>g(n)$
(resp. $g(1)<g(n)$). It is evident that
$S^{<}_{n}=S^{>}_{n}w_{n}$, $S^{>}_{n}=S^{<}_{n}w_{n}$, where $w_{n}=n n-1
\cdots 2 1$.

\bthm{PROPOSITION 2.2.11.}  The diagonal matrices $\Lambda^{\nu }(g)$ are
real and satisfy the following recurrences:\newline
i)  $\dsty\Lambda^{\nu }(g)=(-1)^{n-1}|Q^\nu(gw_{n})|^{2}\Lambda^{\nu
}(gw_{n})$, if $g \in S^{>}_{n}$\newline ii)
$\dsty\Lambda^{\nu}(g)=\Lambda^{\nu}_{[1..n]}(g)=
\frac{1}{\bbox^{\nu}_{[1..n]}}\sum_{\emptyset \neq J\subseteq \{
1,2,\dots ,n-1\} , g\in S_{J}}(-1)^{|J|+1}\Lambda^{\nu}_{J}(g)$, if
$g\in S^{<}_{n}$\newline ii') $\dsty\Lambda^{\nu
}(g)=\frac{1}{\bbox^{\nu}_{[1..n]}} \sum_{g=g^{'}g^{''}\in S_{k}\times
S_{n-k}, 1\leq k\leq n-1}(Q^{\nu}_{[1..k]})^{[g(1)<g(k)]} \Lambda^{\nu
}_{[1..k]}(g^{'})\Lambda^{\nu }_{[k+1..n]}(g^{''})$, \newline if $g\in
S^{<}_{n}$. In particular, $[A^{(\nu )}]^{-1}(g)=[A^{(\nu
)}]^{-1}(gw_{n})=0$ if both $g$ and $gw_{n}$ are not splittable, i.e. if
the minimal Young subgroup containing g (resp. $gw_{n}$) is equal to
$S_{n}$. \ethm

{\bf Proof.}Substituting the formula $(I+(-1)^{n}\hat{R}_{\nu }(w_{n}))
^{-1}=\frac{1}{\bbox^{\nu}_{[1..n]}}(I-(-1)^{n}\hat{R}_{\nu }(w_{n}))$
into formula for
$[A^{(\nu )}]^{-1}$ in Prop.2.2.4 we see immediately that for $g\in S^{<}_{n}$
$$
[A^{(\nu )}]^{-1}(g)=\sum_{\emptyset \neq J\subseteq \{1, 2, \dots, n-1\}}
(-1)^{|J|+1}[A^{(\nu )}_{J}]^{-1}(g)\frac{1}{\bbox^{\nu}_{[1..n]}}\eqno (*)
$$

Then for $g\in S^{>}_{n}$,we again use Prop.2.2.4 and Property 3.ii) in 1.8
The property ii) is immediate from (*) because
$[A^{\nu}_{J}]^{-1}(g)\neq 0 \Rightarrow g\in S_{J}$.
To prove ii') one can use the following lemma which we are going to
state without proof.~\qed

\bthm{LEMMA 2.2.12. (Short recursion for the inverse of $A^{(\nu )}$):} We
have

$$[A^{(\nu )}]^{-1}=(\sum_{k=1}^{n-1}(-1)^{k-1}[A^{(\nu )}_{\{ k\} }]
^{-1}\hat{R}_{\nu }(w_{[1..k]})(I+(-1)^{n}\hat{R}_{\nu }(w_{n})
)^{-1}$$
where $A^{(\nu )}_{\{ k\} }=\hat{R}_{\nu }(S_{k}\times S_{n-k}) (=A^{(\nu )}
_{[1..k]}A^{(\nu )}_{[k+1..n]})$ is just $A^{(\nu )}_{J}$ when $J=\{ k\} $.
\ethm

\bthm{COROLLARY 2.2.13.}  With notations of Remark 2.2.7 and Proposition
2.2.11 we have the following formulas for the diagonal entries of the inverse
of $A^{\nu}$, $\nu $ generic, $|\nu |=n$.
$$[A^{(\nu )}]^{-1}(id)=\sum_{\beta }\frac{(-1)^{b(\beta)+n-1}}
{\bbox^{\nu}_{\beta}}\leqno i)
$$
where the sum is over all generalized bracketings $\beta $ of the word
$12\cdots n$, which have outer brackets.
$$[A^{(\nu )}]^{-1}(id)=\frac{1}{\Box^{\nu}_{[1..n]}}\sum_{\beta }
\frac{Q^{\nu}_{\beta}}{\Box^{\nu}_{\beta}}\leqno i')
$$
where the sum is over all generalized bracketings $\beta $, without
outer brackets, of the word $12\cdots n$,  and where $Q^{\nu}_{\beta}$
is defined, analogously as $\bbox^{\nu}_{\beta}$,
to be the product of $Q^{\nu}_{[a..b]}$ over all bracket pairs in $\beta $.
\ethm
{\bf Proof.} i) follows from Remark 2.2.7 because $\hat{R}_{\nu }$-terms
contribute only to nondiagonal entries.i') follows by iterating
Proposition 2.2.11 $ii)^{'}$ in case $g=id$ and using that $[A^{(\nu
)}]^{-1}(id)=\Lambda^{\nu }(id)A^{(\nu )}(id)= \Lambda^{\nu }(id)Q^{\nu
}(id)=\Lambda^{\nu }(id)$.~\qed
In particular if $I=\{ 1,2\} ,\nu_{1}=\nu_{2}=1$, we have
$\Lambda^{12}(id)=[A^{12}]^{-1}(id)=\frac{1}{\bbox_{\{ 1,2\} }}$.

In Ex.1.6.3 ($I=\{ 1,2,3\} ,\nu_{1}=\nu_{2}=\nu_{3}=1$)we have
 \begin{eqnarray*}
\Lambda^{123}(id)=[A^{123}]^{-1}(id)
                                        &=&\frac{-1}{\bbox_{123 }}+\frac{1}
                                    {\bbox_{12}\bbox_{123}}+\frac{1}{\bbox_{23}\bbox_{123}}\\
                                     &=&\frac{1}{\bbox_{123}}(1+\frac{Q_{12}}{\bbox_{12}}
                                 +\frac{Q_{23}}{\bbox_{23}})\\
\end{eqnarray*}
\def\STrut{\vphantom{\frac{Q_{34}}{\bbox_{34}}}}
Similarly for $I=\{ 1,2,3,4\} ,\nu_{1}=\nu_{2}=\nu_{3}=\nu_{4}=1$ we have
\[
\begin{array}{r@{\,}l@{\,}l}
\Lambda^{1234}(id) =& \multicolumn{2}{l}{[A^{1234}]^{-1}(id)}\\
    =& \frac{1}{\bbox_{1234}} \left\{\STrut 1\right. &+\frac{Q_{12}}{\bbox_{12}}+
    \frac{Q_{23}}{\bbox_{23}}+\frac{Q_{34}}{\bbox_{34}}+
    \frac{Q_{12}Q_{34}}{\bbox_{12}\bbox_{34}} \\
     &                                               &+\left.(1+
     \frac{Q_{12}}{\bbox_{12}}+\frac{Q_{23}}{\bbox_{23}})
        \frac{Q_{123}}{\bbox_{123}}+(1+\frac{Q_{23}}{\bbox_{23}}+
        \frac{Q_{34}}{\bbox_{34}})
    \frac{Q_{234}}{\bbox_{234}}\right\}
\end{array}
 \]
(Here we abbreviated $Q_{\{ 1,2\} }, Q_{\{
2,3,4\} }$ to $Q_{12},Q_{234}$ etc.). If we take all $q_{ij}=q$ (Zagier's
case), then we obtain easily that
 $$[A_{3}(q)]^{-1}(id)=\frac{1+q^{2}}{(1-q^{2})(1-q^{6})}I,\quad
[A_{4}(q)]^{-1}(id)=\frac{1+2q^{2}+q^{4}+2q^{6}+q^{8}}{(1-q^{2})
(1-q^{6})(1-q^{12})}I
$$
what agrees with Zagier's computations.
\rem{REMARK 2.2.14.} The formula i') in Corollary 2.2.13 can be
interpreted also as a regular language expression for closed walks in
the weighted digraph (a Markov chain) ${\cal D}^{\nu }$ on the symmetric
group $S_{n}$ where the adjacency matrix $A({\cal D}^{\nu })$ is given
by nondiagonal entries of $A^{(\nu )}$ multiplied by -1, i.e. $A({\cal
D}^{\nu })=-(A^{(\nu )}-I)$. Then the walk generating matrix function of
${\cal D}^{\nu }$ is nothing but the inverse of $A^{(\nu )}$ because
$W({\cal D}^{\nu })=(I-A({\cal D}^{\nu })) ^{-1}=[A^{(\nu )}]^{-1}$. For
example, we have
\noindent
\[
\begin{array}{lll}
W({\cal D}^{123})_{closed}&=&[A^{123}]^{-1}(id)=Q_{\{1,2,3\} }^
{\hphantom{[1..2]}*} (I+Q_{\{ 1,2\}}^{\hphantom{[1..2]}+}+
Q_{\{ 2,3\} }^{\hphantom{[1..2]}+})\\ W({\cal D}^{1234})_{closed}
     &=&  [A^{1234}]^{-1}(id)=Q^{\hphantom{[1..2]}*}_{[1..4]}
\left\{1+Q^{\hphantom{[1..2]}+}_{[1..2]}+
                        Q^{\hphantom{[1..2]}+}_{[2..3]}+
 Q^{\hphantom{[1..2]}+}_{[3..4]}\right. +\\
\multicolumn{3}{r}{Q^{\hphantom{[1..2]}+}_{[1..2]}Q^{\hphantom{[1..2]}+}_{[3..4]}
     +\left.(1+Q^{\hphantom{[1..2]}+}_{[1..2]}+ Q^{\hphantom{[1..2]}+}_
     {[2..3]})Q^{\hphantom{[1..2]}+}_{[1..3]}+(1+Q^{\hphantom{[1..2]}+}_{[2..3]}
 +Q^{\hphantom{[1..2]}+}_{[3..4]})Q^{\hphantom{[1..2]}+}_{[2..4]}\right\}}
\end{array}
\]
in the familiar formal language notation
($x^{*}=\frac{1}{1-x}, x^{+}= \frac{x}{1-x}$).

Now we turn our attention to computing a general entry of the inverse  of
$A^{\nu }$, $\nu $ generic, $|\nu |=n$.

Let $g\in S_{n}^{<}$(i.e. $ g(1)<g(n)$) be given. Let $J(g)=\{
j_{1}<j_{2}< \cdots <j_{n(g)-1}\} \subset \{ 1,2,\dots ,n-1\} $ be the
label of the minimal Young subgroup of $S_{n}$ containing g. It is clear
that $J(g)$ can be given explicitly as $J(g)=\{ 1\leq j\leq n-1 |
g(1)+g(2)+\cdots +g(j)=1+2+\cdots +j\} $ Then by $\sigma
(g)=J_{1}J_{2}\cdots J_{n(g)} \in \Sigma_{n}$ we denote the subdivision
associated to $J(g)$ i.e
$$J_{1}=J_{1}(g):=[1..j_{1}], J_{2}=J_{2}(g):=[j_{1}+1..j_{2}],\cdots ,
J_{n(g)}:=J_{n(g)}(g)=[j_{n(g)-1}+1..n]
$$
and by $g=g_{1}g_{2}\cdots g_{n(g)}$ we denote the corresponding
factorization of $g$ with $g_{k}\in S_{J_{k}(g)}, 1\leq k\leq n(g)$. By
noting that $g\in S_{J}\Leftrightarrow J\subseteq J(g)$, we can rewrite
the formula Prop.2.2.11 ii) as follows
 \begin{eqnarray*}
\Lambda^{\nu}(g)&=&\Lambda^{\nu}_{[1..n]}(g)=\frac{1}{\bbox^{\nu}_{[1..n]}}
\sum_{\emptyset \neq J\subseteq J(g)}(-1)^{|J|+1}\Lambda^{\nu}_{J}(g)\\
&=&\frac{1}{\bbox^{\nu}_{[1..n]}}\sum_{\emptyset \neq K\subseteq \{ 1,2,\dots
,n(g)-1\} }(-1)^{|K|+1}\Lambda^{\nu}_{J(K)}(g)
\end{eqnarray*}
where $J(K):=\{ j_{k}| k\in K\} \subseteq \{ 1,2,\dots ,n-1\} $.
(Note that if $J(g)=\emptyset $ ($\Rightarrow g$ and $gw_{n}$ are not
splittable), then $\Lambda^{\nu}(g)=0$ by this formula too.)
In terms of subdivisions this can be viewed as a recursion formula:
$$\Lambda_{[1..n]}^{\nu }(g)=\frac{1}{\bbox^\nu_{[1..n]}}\sum_{\tau =
K_{1}K_{2}\cdots K_{l}\in \Sigma_{n(g)}, l\geq 2}(-1)^{l}\Lambda_{I(K_{1})}
^{\nu }(g_{K_{1}})\cdots \Lambda_{I(K_{l})}^{\nu }(g_{K_{l}})\eqno (*)
$$
where $I(K_{s}):=\bigcup_{k\in K_{s}}J_{k}(g)$, $g_{K_{s}}:=\prod_{k\in
K_{s}}g_{k}, s=1,...,l$.

By iterating this recursion formula (*) (as in Theorem 2.2.6, Remark 2.2.7,
Corollary 2.2.13) we obtain
$$\Lambda_{[1..n]}^{\nu }(g)=(\sum_{\beta }(-1)^{b(\beta )+n(g)-1}\tilde{\Psi }
_{\beta })\Lambda_{J_{1}(g)}^{\nu }(g_{1})\cdots \Lambda^{\nu }
_{J_{n(g)}(g)}(g_{n(g)})\eqno(**)$$
where $\beta $ run over all generalized bracketings of the word $12\cdots
n(g)$ which have outer brackets and where each bracket pair $[a..b]$, $1
\leq a<b\leq n(g)$, we set
$$\tilde{\Psi}_{[a..b]}:=\frac{1}{\bbox_{J_{a}\bigcup J_{a+1}\bigcup
\cdots \bigcup J_{b}}}
$$
($b(\beta):=$number of bracket pairs in $\beta $). Thus the expression in the
parentheses can be viewed as a {\em "thickened" identity coefficient}
$$\Lambda^{12\cdots n(g)}(id)|^{\nu }_{1\rightarrow J_{1}, 2\rightarrow
J_{2},\cdots ,n(g)\rightarrow J_{n(g)}}$$
which we shall denote by
$$\Lambda^{\nu }_{\sigma (g)}=\Lambda^{\nu}
_{J_{1}(g)J_{2}(g)\cdots J_{n(g)}(g)}:=\Lambda^{12\cdots n(g)}(id)
|_{1\rightarrow J_{1},2\rightarrow J_{2},\dots ,n(g)\rightarrow J_{n(g)}}.$$
(In particular we can now write $\Lambda^{\nu}_{[1..n]}(id)$
also as $\Lambda^{\nu}_{[1][2]\cdots [n]}$).

As an example for this notation we take $g=4 1 3 2 5 7 8 6$. Then
$\sigma (g)=[1..4][5][6..8]$ i.e $J_{1}(g)=[1..4], J_{2}(g)=[5],
J_{3}(g)=[6..8]$. So
$$\Lambda^{\nu}_{[1..4][5][6..8]}=\Lambda^{123}(id)|^{\nu}
_{1\rightarrow [1..4], 2\rightarrow [5], 3\rightarrow [6..8]}
=\frac{1}{\bbox^{\nu}_{[1..8]}}(-1+\frac{1}{\bbox^{\nu}_{[1..5]}}
+\frac{1}{\bbox^{\nu}_{[5..8]}})$$(c.f. Corollary 2.2.13).

Now we have one more observation concerning the formula ($**$). To each
nonzero factor $\Lambda^{\nu}_{J_{k}(g)}(g_{k}), 1\leq k\leq n(g)$ in
($**$) we can apply Prop. 2.2.11 i) because $g_{k}$, being a minimal
Young factor
of $g$, is not splittable and hence \newline
$g_{k}(j_{k-1}+1)>g_{k}(j_{k})$ (otherwise $g_{k}w_{J_{k}}$ would also
be nonsplittable $\Rightarrow \Lambda^{\nu}_{J_{k}(g)}(g_{k})=0$)
$$\Lambda^{\nu}_{J_{k}(g)}(g_{k})=(-1)^{|J_{k}(g)|-1}|Q^{\nu}(g_{k}
w_{J_{k}(g)})|^{2}\Lambda^{\nu}_{J_{k}(g)}(g_{k}w_{J_{k}(g)})$$
Substituting this into ($**$) we obtain the following algorithm
for computing the diagonal matrices $\Lambda^{\nu}(g)$ describing
the inverse of $A^{(\nu )}$(recall $[A^{(\nu )}]^{-1}=
\sum_{g\in S_{n}}\Lambda^{\nu}(g)\hat{R}(g)$).
\bthm{PROPOSITION 2.2.15.(An algorithm for $\Lambda^{\nu}(g)$, $\nu $
generic, $|\nu |=n$).}For $g\in S_{n}$ we have
$$\Lambda^{\nu}_{[1..n]}(g)=(-1)^{n-n(g)}\Lambda^{\nu}_{\sigma (g)}
|Q^{\nu}(g^{'})|^{2}\Lambda^{\nu}_{J(g)}(g^{'})$$
where $g^{'}:=gw_{J(g)}$ ($w_{J(g)}=$ the maximal element in the
minimal Young subgroup $S_{J(g)}$ containing $g$). Similar statement holds
true if we replace $[1..n]$ by any interval $[a..b], 1\leq a\leq b\leq n$.
\ethm

{\bf Proof.}  If $g(1)<g(n)$ this is what we get from ($**$).If
$g(1)>g(n)$, then $J(g)=\emptyset , S_{J(g)}=S_{n}, w_{J(g)}= n n-1
\dots 2 1=w_{n}, n(g)=1, \sigma (g)=[1..n], \Lambda^{\nu}
_{\sigma(g)}=\Lambda^{1}(id)|_{1\rightarrow [1..n]}=I, g^{'}=gw_{J(g)}=
gw_{n}$, so what we needed to prove is just the claim in Prop.2.2.11
i).~\qed To ilustrate this algorithm we take (again!) $g=41325786$ ($\nu
$ generic weight, $|\nu |=8$) for which $J(g)=\{ 4,5\} ,
J_{1}(g)=[1..4], J_{2}(g)=[5], J_{3}(g)=[6..8], n(g)=3, n=8, w_{J(g)}
=43215876, g^{'}=gw_{J(g)}=23145687, Q^{\nu}(g^{'})=Q^{\nu}_{1,2}
Q^{\nu}_{1,3}Q^{\nu}_{7,8},$\newline $
|Q^{\nu}(g^{'})|^{2}=Q^{\nu}(g^{'}) Q^{\nu}(g^{'})^{*}=Q^{\nu}_{\{ 1,2\}
}Q^{\nu}_{\{ 1,3\} }Q^{\nu}_{\{7,8\} }$. Then the first step of our
algorithm gives
\begin{eqnarray*}
&&\Lambda^{\nu}_{[1..8]}(g)=\Lambda^{\nu}_{[1..8]}(41325786)=\\
&&=(-1)^{8-3}\Lambda^{\nu}_{[1..4][5][6..8]}Q^{\nu}_{\{ 1,2\} }
Q^{\nu}_{\{ 1,3\} }Q^{\nu}_{\{ 7,8\} }\Lambda^{\nu}_{[1..4]}(2314)
\Lambda^{\nu}_{[5]}(5)\Lambda^{\nu}_{[6..8]}(687).
\end{eqnarray*}
 In the second step of our algorithm we compute\newline
$\Lambda^{\nu}_{[1..4]}(2314)=(-1)^{4-2}\Lambda^{\nu}_{[1..3][4]}
Q^{\nu}_{\{ 2,3\}
}\Lambda^{\nu}_{[1..3]}(132)\Lambda^{\nu}_{[4]}(4)$\newline
$\Lambda^{\nu}_{[6..8]}(687)=(-1)^{3-2}\Lambda^{\nu}_{[6][7..8]}
\Lambda^{\nu}_{[6]}(6)\Lambda^{\nu}_{[7..8]}(78)$\newline In the third
(last) step we need only to compute
 $$\Lambda^{\nu}_{[1..3]}(132)=(-1)^{3-2}\Lambda^{\nu}_{[1][2..3]}
\Lambda^{\nu}_{[1]}(1)\Lambda^{\nu}_{[2..3]}(23).
$$
Since $\Lambda^{\nu}_{[7..8]}(78)=\Lambda^{\nu}_{[7][8]}, \Lambda^{\nu}
_{[2..3]}(23)=\Lambda^{\nu}_{[2][3]}, (Q^{\nu}_{\{ 1,2\} }Q^{\nu}_{\{
1,3\} }) Q^{\nu}_{\{ 2,3\} }=Q^{\nu}_{[1..3]},
\Lambda^{\nu}_{[1]}(1)=\cdots =\Lambda^{\nu}_{[8]}(8)=I$, we finally
obtain
$$\Lambda^{\nu}_{[1..8]}(41325786)=-\Lambda^{\nu}_{[1..4][5][6..8]}
\Lambda^{\nu}_{[1..3][4]}\Lambda^{\nu}_{[1][2..3]}\Lambda^{\nu}
_{[2][3]}\Lambda^{\nu}_{[6][7..8]}\Lambda^{\nu}_{[7][8]}Q^{\nu}_{[1..3]}
Q^{\nu}_{[7..8]}.$$

As a general example we take $g=w_{J}$ where $J=\{ j_{1}<\cdots <j_{l-1}\} $
is an arbitrary subset of $\{ 1,2,\dots ,n-1\} $. Here $n(g)=l$ and $g^{'}
=id$, so by one application of our algorithm we obtain
$$\Lambda^{\nu}_{[1..n]}(w_{J})=(-1)^{n-l}\Lambda^{\nu}_{J_{1}J_{2}
\cdots J_{l}}\Lambda^{\nu}_{J_{1}}(id)\Lambda^{\nu}_{J_{2}}(id)\cdots
\Lambda^{\nu}_{J_{l}}(id)$$
where $J_{1}=[1..j_{1}], J_{2}=[j_{1}+1..j_{2}], \dots ,J_{l}=[j_{l-1}+1..n]$.In particular for $n=8$, $J=\{ 4\} $ we obtain
\begin{eqnarray*}
\Lambda^{\nu}_{[1..8]}(43218765) &=&(-1)^{8-2}\Lambda^{\nu}_{[1..4][5..8]}
\Lambda^{\nu}_{[1..4]}(1234)\Lambda^{\nu}_{[5..8]}(5678)\\
&=&\frac{1}{\bbox^{\nu}_{[1..8]}}\Lambda^{\nu}_{[1][2][3][4]}\Lambda^{\nu}
_{[5][6][7][8]}
\end{eqnarray*}
In Zagier's case, when all $q_{ij}=q$, we would then have (c.f. Examples
to Cor. 2.2.13)
$$\Lambda^{\nu}_{[1..8]}(43218765)=\frac{1}{1-q^{7\cdot 8}}\frac{
(1+2q^{2}+q^{4}+2q^{6}+q^{8})^{2}}{(1-q^{1\cdot 2})^{2}(1-q^{2\cdot 3})^{2}
(1-q^{3\cdot 4})^{2}}I$$
But the denominator $D_{8}$ of this expression does not divides Zagier's
$\triangle_{8}=(1-q^{2\cdot 1})(1-q^{3\cdot 2})(1-q^{4\cdot 3})(1-q^{5\cdot 4})
(1-q^{6\cdot 5})(1-q^{7\cdot 6})(1-q^{8\cdot 7})$. Namely $\triangle_{8}
/D_{8}=(1-q^{4\cdot 5})(1-q^{5\cdot 6})(1-q^{6\cdot 7})/(1-q^{1\cdot 2})
(1-q^{2\cdot 3})(1-q^{3\cdot 4})$ is not a polynomial due to the factor
$1-q^{2}+q^{4}$ in the denominator. This computation shows
that the original Zagier's conjecture (c.f. Remark 2.2.9) fails for $n=8$.

Now we return back to our algorithm. We shall show now that it is somewhat
better to combine two steps of our algorithm into one step. This can be
observed already in our illustrative example ($g=41325786$) where after the
second step the "unrelated factors" $Q^{\nu}_{\{ 1,2\} }$ and $Q^{\nu}_{\{
1,3\} }$ from the first step were completed, with the factor $Q^{\nu}_{\{
2,3\} }$, into a "nicer" term $Q^{\nu}_{[1..3]}$ having a contiguous indexing set.
Fortunately this holds in general, but first we need more notations to state
the results. To each permutation $g\in S_{n}$ we can associate a sequence of permutations
$g, g^{'}, g^{''}, \dots $, where $g^{(k+1)}$ is obtained from $g^{(k)}$
by reversing all minimal Young factors in $g^{(k)}$ i.e $g^{'}=gw_{J(g)},
g^{''}=gw_{J(g^{'})},\dots ,g^{(k+1)}=(g^{(k)})^{'}=g^{(k)}w_{J(g^{(k)})},
\dots $.We shall call this sequence a {\em Young sequence} of $g$. Further we call $g$ {\em tree-like} if $g^{(k)}=id$ for some $k$, and by {\em depth
of} $g$ we call the minimal such $k$.
Besides the notation $\Lambda^{\nu}_{\sigma (g)}=\Lambda^{\nu}_{J_{1}(g)
J_{2}(g)\cdots J_{n(g)}(g)}$, where $\sigma (g)=J_{1}(g)\cdots J_{n(g)}(g)$
is the subdivision of $\{ 1,2,\dots ,n\} $ associated to the minimal
Young subgroup $S_{J(g)}$ containing $g$ we need a relative version
$\Lambda^{\nu}_{\sigma (g^{'}):\sigma (g)}$ which we define by
$$\Lambda^{\nu}_{\sigma (g^{'}):\sigma (g)}:=\Lambda^{\nu}
_{\sigma (g^{'}|J_{1}(g))}\Lambda^{\nu}_{\sigma (g^{'}|J_{2}(g))}\cdots
\Lambda^{\nu}_{\sigma (g^{'}|J_{n(g)}(g))}$$
For example when $g=41325786 (\Rightarrow g^{'}=23145687)$, $J_{1}(g)
=[1..4], J_{2}(g)=[5], J_{3}(g)=[6..8]$, we have
$\Lambda^{\nu}_{\sigma (g^{'}):\sigma (g)}=\Lambda^{\nu}_{[123][4]}
\Lambda^{\nu}_{[5]}\Lambda^{\nu}_{[6][7..8]}$.
Also, besides the notation, for $T\subseteq \{ 1,2,\dots ,n\} ,
Q^{\nu}_{T}=\prod_{a,b\in T,a\neq b}Q^{\nu}_{a,b}$
(introduced in 1.8), we define for any
subdivision $\sigma =J_{1}J_{2}\cdots J_{l}$ of $\{ 1,2,\dots ,n\} $:
$$
Q^{\nu}_{\sigma }:=Q^{\nu}_{J_{1}}Q^{\nu}_{J_{2}}\cdots Q^{\nu}_{J_{l}}
$$
For example:
$Q^{\nu}_{[1..3][4][5][6][7..8]}=Q^{\nu}_{[1..3]}Q^{\nu}_{[4]}Q^{\nu}_{[5]}
Q^{\nu}_{[6]}Q^{\nu}_{[7..8]}=Q^{\nu}_{[1..3]}Q^{\nu}_{[7..8]}.$
\bthm{PROPOSITION 2.2.16.(Fast algorithm for $\Lambda^{\nu}(g)$, $\nu $ generic,
$|\nu |=n$):} With the notations above we have
$$\Lambda^{\nu}_{[1..n]}(g)=(-1)^{n(g)+n(g^{'})}\Lambda^{\nu}_{\sigma (g)}
\Lambda^{\nu}_{\sigma (g^{'}):\sigma (g)}Q^{\nu}_{\sigma (g^{'})}\Lambda^{\nu}
_{J(g^{'})}(g^{''})$$
($n(g)=$ the number of minimal Young factors of $g$)
\ethm
 {\bf Proof.} By applying twice the algorithm in Proposition 2.2.15.~\qed
 Now we shall state our principal result concerning the inversion of
 matrices $A^{(\nu )}$ of the sesquilinear form $(\ ,\ )_{\bf q}$,
 defined in 1.3, on the generic weight space ${\bf f}_{\nu}, |\nu |=n$.

 \bthm{THEOREM 2.2.17. [INVERSE MATRIX ENTRIES]} Let $\nu $ be a
 generic weight, $|\nu |=n$. For the coeficients $\Lambda^{\nu}(g)$ in
 the expansion $$[A^{(\nu )}]^{-1}=\sum_{g\in
 S_{n}}\Lambda^{\nu}(g)\hat{R}_{\nu}(g)$$ we have, with the notations
 above, the following formulas:\newline i) If $g\in S_{n}$ is a
 tree-like permutation of depth $d$, then
 $$\Lambda^{\nu}(g)=(-1)^{N}\Lambda^{\nu}_{\sigma (g)}
 \Lambda^{\nu}_{\sigma (g^{'}):\sigma (g)}\Lambda^{\nu}_{\sigma (g^{''})
 :\sigma (g^{'})}\cdots \Lambda^{\nu}_{\sigma (g^{(d)}):\sigma
 (g^{(d-1)})} Q^{\nu}_{\sigma (g^{'})}Q^{\nu}_{\sigma (g^{'''})}\cdots
 Q^{\nu}_{\sigma (g^{(d^{'})})}$$ where $\dsty
 N=N(g):=\sum_{k=0}^{d}\sum_{I\in \sigma (g^{(k)})}(Card\ I-1), \ d^{'}=
 2\left\lfloor{(d-1)/2}\right\rfloor+1$\newline ii) If $g\in S_{n}$
 is not tree-like, then $\Lambda^{\nu}(g)=0$. \ethm
{\bf Proof.} i) follows by iterating our fast algorithm (of
Prop.2.2.16).ii) If $g$ is not tree-like then in the Young sequence of
$g$ we encounter some Young factor which together with its reverse is
not splittable, but then the corresponding
$\Lambda^{\nu}_{[..]}(\hbox{the factor})=0$ (c.f. Prop.2.2.11), hence
$\Lambda^{\nu}(g)=0$.~\qed

Now we give explicit formulas for the inverses of $A^{123}$ and
$A^{1234}$:We have 
\begin{eqnarray*}
[A^{123}]^{-1}&=&\frac{1}{\bbox_{[1..3]}}\{ \frac{I-Q_{[1..2]}Q_{[2..3]}}
{\bbox_{[1..2]}\bbox_{[2..3]}}(\hat{R}(123)+\hat{R}(321))-\\
&-&\frac{1}{\bbox_{[1..2]}}(\hat{R}(213)+Q_{[1..2]}\hat{R}(312))
-\frac{1}{\bbox_{[2..3]}}(\hat{R}(132)+Q_{2..3]}\hat{R}(231))\}.\\{}
[A^{1234}]^{-1}&=&\Lambda^{1234}(id)\hat{R}(1234)+\frac{1}{\bbox_{1234}}
\{ -\frac{I-Q_{123}Q_{34}}{\bbox_{12}\bbox_{123}\bbox_{34}}\hat{R}(2134)\\
&-&\frac{I-Q_{123}Q_{234}}{\bbox_{23}\bbox_{123}\bbox_{234}}\hat{R}(1324)
-\frac{I-Q_{12}Q_{234}}{\bbox_{12}\bbox_{34}\bbox_{234}}\hat{R}(1243)
+\frac{1}{\bbox_{12}\bbox_{34}}\hat{R}(2143)+\\
&+&\frac{I-Q_{12}Q_{23}}{\bbox_{12}\bbox_{23}\bbox_{123}}\hat{R}(3214)-
\frac{Q_{12}}{\bbox_{12}\bbox_{123}}\hat{R}(3124)
-\frac{Q_{23}}{\bbox_{23}\bbox_{123}}\hat{R}(2314)\\
&+&\frac{I-Q_{23}Q_{34}}{\bbox_{23}\bbox_{34}\bbox_{234}}\hat{R}(1432)
-\frac{Q_{23}}{\bbox_{23}\bbox_{234}}\hat{R}(1423)
-\frac{Q_{34}}{\bbox_{34}\bbox_{234}}\hat{R}(1342)\} \\
&+&\hbox{(eleven terms obtained by multiplying with $-\hat{R}(4321)$)}.
\end{eqnarray*}
where $\Lambda^{123}(id)$ and $\Lambda^{1234}(id)$ are given as examples
 illustrating Corollary 2.2.13.(Here we abbreviated $Q_{[1..2]},
 Q_{[2..4]}$ to $Q_{12}$ (not to be confused with $Q_{1,2}$), $Q_{234}$
 etc.). Note that $A^{1234}$ is a $24\times 24$ symbolic matrix so the
 inversion of such a matrix by standard methods on a computer is almost
 impossible (the output may contain  huge number of pages of messy
 expressions!).
\rem{REMARK 2.2.18.} By using our reduction to the generic case
formula 1.7.1) $[A^{(\nu)}]_{\bf ij}^{-1} = \sum_{h\in H} [\tilde
A^{(\tilde\nu)}]_{{\bf\tilde i}, h{\bf\tilde j}}^{-1}$ we can
write also formulas for the inverse matrix entries in the case of
degenerate weights $\nu$. E.g. for the inverse of $A^{113}$ (see Example
1.6.4) one gets
 $$[A^{113}]^{-1}=\frac1\Delta\left(\begin{array}{ccc}
1 & -(1+q_{11})q_{13}&q_{11}q_{13}^2\\
-q_{31}(1+q_{11}) & (1+q_{11})(1+q_{13}q_{31}) & -(1+q_{11})q_{13}\\
q_{13}^2q_{11} & -q_{31}(1+q_{11}) & 1
\end{array}\right)$$
where $\Delta=(1+q_{11})(1-q_{13}q_{31})(1-q_{11}q_{13}q_{31}).$
\section{Applications} 
\subsection{Quantum bilinear form of the discriminant arrangement of
hyperplanes} 
Here we briefly recall the definition of the quantum bilinear form
in case of the configuration ${\cal A}_{n}$ of diagonal hyperplanes
$H_{ij}=H_{ij}^{n}:x_{i}=x_{j}, 1\leq i<j\leq n$ in ${\bf R}^{n}$ (for
general case see [Var]). This arrangement ${\cal A}_{n}$ is also called
the {\em discriminant arrangement} of hyperplanes in ${\bf R}^{n}$.
The {\em domains} of ${\cal A}_{n}$ (i.e connected components of the complement
of the union of hyperplanes in ${\cal A}_{n}$) are clearly of the form $$P_{\pi }=\{ x\in {\bf R}^{n}| x_{\pi (1)}<x_{\pi (2)}<\cdots <x_{\pi (n)}\}
,\pi \in S_{n}$$
Let $a(H_{ij}^{n})=q_{ij}$ be the {\em weight} of the hyperplane $H_{ij}\in {\cal A}_{n}$, where $q_{ij}$ are given real numbers, $1\leq i<j\leq n$.
Then the {\em quantum bilinear form} $B_{n}$ of ${\cal A}_{n}$ is defined on the
free vector space $M_{n}=M_{{\cal A}_{n}}$, generated by the domains of
${\cal A}_{n}$, by $$B_{n}(P_{\pi },P_{\tau })=\prod a(H)$$
where the product is taken over all hyperplanes $H\in {\cal A}_{n}$
which separate $P_{\pi }$ from $P_{\tau}$.
\bthm{PROPOSITION 3.1.1.} We have
$$B_{n}(P_{\pi },P_{\tau })=\prod_{(a,b)\in I(\pi^{-1})\triangle I(\tau^{-1})}
q_{ab}$$
where $I(\sigma )=\{ (a,b)| a<b, \sigma (a)>\sigma (b)\} $ denotes the set of
inversions of $\sigma \in S_{n}$ and $X\triangle Y=(X\setminus Y)\bigcup
(Y\setminus X)$ denotes the symmetric difference of sets $X$ and $Y$.
\ethm
\bthm{COROLLARY 3.1.2.} The matrix of the quantum bilinear form $B_{n}$ of
the discriminant arrangements ${\cal A}_{n}=\{ H_{ij}\} $ of hyperplanes in
${\bf R}^{n}$ coincides with the matrix $A^{12\cdots n}=A^{12\cdots n}
({\bf q})$ of the form $(\ ,\ )_{{\bf q}}$ (defined in 1.3), restricted
to the generic weight
space ${\bf f}_{\nu}$, where $I=\{ 1,2,\dots ,n\} ,\nu_{1}=\nu_{2}=\cdots
=\nu_{n}=1$ and where ${\bf q}=\{ q_{ij}\in {\bf R}, 1\leq i, j\leq n,
q_{ij}=q_{ji}\} , q_{ij}=$ the weight of $H_{ij}$ for $1\leq i<j\leq n$.
\ethm
    This Corollary enables us to translate all our results concerning matrices
$A^{\nu}, \nu =$ generic, $|\nu |=n$ into results about the quantum bilinear
form $B_{n}$. As an example we shall reinterpret our determinantal formula
given in Theorem 1.9.2.
\bthm{THEOREM 3.1.3.} The determinant of the quantum bilinear form $B_{n}$
of the discriminant arrangement ${\cal A}_{n}$ is given by the formula
$$\det B_{n}=\prod_{L\in {\cal E}^{'}({\cal A}_{n})}(1-a(L)^{2})^{l(L)}$$
where for $L=\{ x_{i_{1}}=x_{i_{2}}=\cdots =x_{i_{k}}\} \in {\cal A}_{n,k}
\subset {\cal E}^{'}({\cal A}_{n})$ we have
$$a(L)=\prod_{1\leq a<b\leq k}q_{i_{a}i_{b}},\,\, l(L)=(k-2)!(n-k+1)!$$
\ethm
Note that our formula for $\det B_{n}$ is more explicit then Varchenko's
formula, and in particular we conclude that the multiplicity
$l(L)=0$ for all $L\in {\cal E}({\cal A}_{n})\setminus{\cal E}^{'}({\cal
A}_{n})$.
Note added in proof.After receiving a new book by Varchenko [Multidimensional
hypergeometric functions and representation theory of Lie algebras and quantum
groups,World Scientific (1995)] we found in it a result ( Theorem 3.11.2.-proved by
different techniques) equivalent to our Theorem 3.1.3 ,but there seems
to be no results equivalent to our inversion formulas applied to discriminant
arrangements.
\subsection{Quantum groups} 
 We shall use the notations from  [SVa]. Our Theorem 1.9.2 implies the following
 \bthm{THEOREM 3.2.1.} The determinant of the contravariant form $S$ on the
weight space $(U_{q}{\goth n}_{-})_{(1,1,\dots ,1)}$ is given by the following
formula
\begin{eqnarray*}
\det S|_{(U_{q}{\bf n}_{-})_{(1,1\dots ,1)}}=\\
=q^{-\frac{n!}{4}\sum_{1\leq k<l\leq
n}b_{kl}}\prod_{m=2}^{n}\prod_{1\leq i_{1}<\cdots <i_{m}\leq n}(1-q^{\sum_{
1\leq k<l\leq m}b_{i_{k}i_{l}}})^{(m-2)!(n-m+1)!}\\
=\prod_{m=2}^{n}\prod_{1\leq i_{1}<\cdots <i_{m}\leq n}(q^{-\frac{1}{2}
\sum_{1\leq k<l\leq m}b_{i_{k}i_{l}}}-q^{\frac{1}{2}\sum_{1\leq k<l\leq m}
b_{i_{k}i_{l}}})^{(m-2)!(n-m+1)!}
\end{eqnarray*}
\ethm
{\bf Proof.} By factoring out from the matrix $S(f_{I},f_{J})$
the factor $q^{-\frac{1}{4}\sum_{1\leq k<l\leq n}b_{kl}}$ we get a matrix which
(up to permutation of rows and columns) coincides with the matrix $A^{12\cdots
n}({\bf q})$, where ${\bf q}=\{ q_{ij}\} $, $q_{ij}:=q^{-\frac{1}{2}b_{ij}}$.
Then we apply Theorem 1.9.2 and the result follows.

\end{document}